\documentclass{emulateapj}

\usepackage{color}

\slugcomment{Accepted for publication in ApJ}

\shorttitle{Studying the WHIM in emission}
\shortauthors{Takei et al.}

\newcommand{\kms}{{\rm km} \ {\rm s}^{-1}}

\newcommand{\nbin}{n_\mathrm{bin}}
\newcommand{\XENIA}{{\it Xenia} }
\newcommand{\DE}{{\Delta E}}
\newcommand{\deltag}{{\delta_g}}
\newcommand{\rhogas}{\rho_\mathrm{gas}}
\newcommand{\phot}{ph s$^{-1}$ cm$^{-2}$ sr$^{-1}$}
\newcommand{\photkeV}{ph s$^{-1}$ cm$^{-2}$ sr$^{-1}$ keV$^{-1}$}
\newcommand{\photeV}{ph s$^{-1}$ cm$^{-2}$ sr$^{-1}$ eV$^{-1}$}
\newcommand{\tcool}{t_\mathrm{cool}}
\newcommand{\tH}{t_\mathrm{H}}

\begin{document}

\title{Studying the Warm-Hot Intergalactic Medium  in Emission}


\author{ 
Y.~Takei\altaffilmark{1}, E.~Ursino\altaffilmark2,
E.~Branchini\altaffilmark2, T.~Ohashi\altaffilmark{3},
H.~Kawahara\altaffilmark{3}, K.~Mitsuda\altaffilmark{1},\\
L.~Piro\altaffilmark4, 
A. ~Corsi\altaffilmark4, L.~Amati\altaffilmark5,
J.W. den Herder\altaffilmark6, M.~Galeazzi\altaffilmark7,
J.~Kaastra\altaffilmark{6,8}, \\
L.~Moscardini\altaffilmark{9,10},
F.~Nicastro\altaffilmark{11},
F.~Paerels\altaffilmark{12},
M.~Roncarelli\altaffilmark{9} and
M.~Viel\altaffilmark {13,14}}

\altaffiltext{{1}}{ Institute of Space and Astronautical Science,
Japan Aerospace Exploration Agency,
3-1-1 Yoshinodai, Chuo-ku, Sagamihara, Kanagawa 252-5210, Japan}
\email{takei@astro.isas.jaxa.jp}
\altaffiltext{2}{Dipartimento di Fisica, Universit\`a degli Studi ``Roma Tre'' via della Vasca Navale 84, I-00146 Roma, Italy}
\altaffiltext{3}{ Department of Physics, School of Science, Tokyo Metropolitan University, 1-1 Minami-Osawa, Hachioji, Tokyo 192-0397, Japan}
\altaffiltext{4}{INAF-Istituto di Astrofisica Spaziale Fisica Cosmica, Via del Fosso del Cavaliere 100, I-00133 Roma, Italy}
\altaffiltext{5}{INAF-Istituto di Astrofisica Spaziale e 
Fisica Cosmica Bologna, via P. Gobetti 101, I-40129
Bologna, Italy}  
\altaffiltext{6}{ SRON Netherlands Institute for Space Research,
Sorbonnelaan 2, 3584 CA Utrecht, The Netherlands}  
\altaffiltext{7}{ Physics Department of University of Miami,
319 Knight Physics Building, Coral Gables, FL 33164, U.S.}
\altaffiltext{{8}}{ Astronomical Institute, University of Utrecht, Postbus 8000, 3508, TA Utrecht, The Netherlands}
\altaffiltext{9}{ Dipartimento di Astronomia, Universit\`a  di Bologna, via Ranzani 1, I-40127 Bologna, Italy}
\altaffiltext{{10}}{ INFN/National Institute for Nuclear Physics,
Sezione di Bologna, Via Berti Pichat 6/2, I-40127, Bologna, Italy}
\altaffiltext{11}{ INAF-Osservatorio Astronomico di Roma,via Frascati 33, I00040 Monteporzio-Catone (RM), Italy}
\altaffiltext{{12}}{ Columbia Astrophysics Laboratory and Department of Astronomy, Columbia University, 550 West 120th Street, New York, NY 10027, U.S.}   
\altaffiltext{{13}}{ INAF-Osservatorio Astronomico di Trieste,  via Tiepolo 11, I-34131 Trieste, Italy}
\altaffiltext{{14}}{ INFN/National Institute for Nuclear Physics, Via Valerio 2, I-34127 Trieste, Italy}

\begin{abstract}

We assess the possibility to detect the warm-hot intergalactic
medium (WHIM) in emission and to
characterize its physical conditions and spatial distribution through
spatially resolved X-ray spectroscopy, in the framework of the recently
proposed {\it DIOS}, {\it EDGE}, {\it Xenia}, and 
{\it ORIGIN} missions, all of which
are  equipped
with microcalorimeter-based detectors.  For this purpose we analyze a
large set of mock emission spectra, extracted from a cosmological
hydrodynamical simulation. 
These mock X-ray spectra are
searched for emission features showing both the \ion{O}{7} K$\alpha$
 triplet and
\ion{O}{8} Ly$\alpha$
 line, which constitute a typical signature of the warm hot gas.
Our analysis shows that 1 Ms long exposures and energy resolution 
of 2.5 eV will allow us to detect about 400 such features per deg$^{2}$
with a significance $\ge 5\sigma$ and reveals that these emission systems
are typically associated with density $\sim$100 above 
the mean.
The temperature can be 
estimated from the line ratio with a precision of $\sim 20$
\%. The combined effect of contamination from other lines, variation in
the level of the continuum, and degradation of the energy resolution
reduces these estimates.  Yet, with an energy resolution of 7 eV and all
these effects taken into account, one still expects about 160 detections
per deg$^2$.  These line systems are sufficient to trace the spatial
distribution of the line-emitting gas, which constitute an additional
information, independent from line statistics, to constrain the poorly
known cosmic chemical enrichment history and the stellar feedback
processes.

\end{abstract}

\keywords{
cosmology: observations ---
intergalactic medium ---
large-scale structure of Universe ---
X-rays: diffuse background
}
\maketitle

\section{Introduction}
\label{sec:introduction}
After more than 10 years, the so-called ``missing 
baryon'' problem is far from being solved.
The original claim of \citet*{1998ApJ...503..518F} that the 
observational census of cosmic baryons in the local Universe falls 
short of the expected cosmological mean 
\citep{1997AJ....114.1330B,1998AJ....115.1725B,2003ApJS..148..175S,2009ApJS..180..330K}
has gained even more statistical significance 
\citep{2004ApJ...616..643F,2005ApJ...624..555D}
and 
it is now generally accepted that a large fraction  of the baryons 
($\sim$40~\%) is not accounted for by observations at $z \sim 0$. This 
fact is even more surprising if one considers the fact that the baryon 
density in the Ly-$\alpha$ forest at redshift $z=$2--3 
\citep{1997ApJ...490..564W,1998ARA&A..36..267R}
accounts for about all expected baryons whereas the nucleons in the 
Ly-$\alpha$ forest at $z \sim 0$
contributes to less than 30~\% of the cosmic mean
\citep{2007ARA&A..45..221B}.

Although  baryons provide a small contribution to the  cosmological 
mass-energy density, they are key players 
in forming stars and galaxies and  determining the cosmochemical evolution of the diffuse gas. In addition, being the only component that interacts
with the electromagnetic radiation, they are the sole probe to investigate  
the physics of the formation and evolution of visible cosmic structures
and infer the co-evolution of the underlying dark matter distribution.

Cosmological numerical simulations have provided a solution to this 
apparent paradox suggesting that missing baryons at $z<1$ are 
distributed in a network of filaments with temperature  
$10^5$--$10^7{\rm~K}$ with density significantly above the 
cosmic mean but below that typically found in virialized structures like
galaxy clusters.  
This phase is called warm-hot intergalactic medium (WHIM).
These results, which constitute one of the 
rare examples in cosmology in which theory leads observations, 
explain why these baryons have escaped observations so far and indicate 
the best strategy to detect them.
Indeed, the WHIM is expected to be so highly ionized that it can only be seen in 
the UV and X-ray bands but the signal should be very weak because of
its relatively low density. 
Therefore it is not surprising that current observations
in the UV and X-ray bands provide weak constraints to the WHIM models.

In the UV band, \ion{O}{6} absorption lines ($\lambda=1032, \ 1038 ~\mathrm{\AA}$)
have been detected along the lines of sight to $\sim 50$ AGN 
\citep[e.g.,][]{2005ApJ...624..555D,2008ApJ...679..194D,2008ApJS..177...39T}.
Their
cumulative number per unit redshift as a function of the line equivalent
width (hereafter  the $dN/dz$ statistics) are
consistent with the predictions of numerical models
\citep[see e.g.,][]{2006ApJ...650..573C,2009ApJ...697..328B}.
However, it is debated how reliably
the observed \ion{O}{6} absorption lines trace the WHIM since their width seems
too small to be produced by warm-hot material.  This means that either
the observed lines trace the local Ly-$\alpha$ forest rather than the
WHIM or that the gas is two-phased: a photo-ionized gas responsible for
the Ly-$\alpha$ absorption and shock-heated WHIM responsible for the \ion{O}{6}
lines. 
Recent numerical simulations 
carried out to investigate the physical properties of
the IGM have helped to clarify the issue
\citep{2009MNRAS.393...99W,2009MNRAS.395.1875O,2010arXiv1009.0261S,2011MNRAS.tmp..165T}.
These simulations have shown that the  thermal state of the IGM at $z\sim 0$
is significantly affected by a number of physical processes like
galactic wind, AGN feedback, metal line cooling,
photoionization by background radiation, sub-resolution
turbulence.
Although it is non trivial to compare the results of different 
works and not all these works seem to converge to the same results,
it is remarkable that 
two of them \citep{2010arXiv1009.0261S,2011MNRAS.tmp..165T} 
indicate that about  a third of the OVI absorbers are photo-ionized
and trace the Ly-$\alpha$ forest at $z\sim 0$ while the rest 
of them trace the WHIM.
An upper limit can nevertheless been worked out: if all gas is
multi-phased, i.e., in the limit in which all the observed \ion{O}{6} lines are
produced by the WHIM, then the \ion{O}{6} can probe no more than 7-10~\% of the
baryons 
\citep{2005ApJ...624..555D,2006ASPC..348..341T,2009AIPC.1135....8D}.
With the inclusion of the thermally broadened Ly$\alpha$ absorbers 
(BLAs) recently observed by
\cite{2010ApJ...710..613D} and
\cite{2010ApJ...712.1443N,2010ApJ...721..960N},
these estimates are boosted up by a factor of $\sim$2, 
i.e., UV observations are capable of probing 
$\sim$20\% of the baryons in the local universe
\citep{2010ApJ...710..613D}.
Nevertheless, whereas \ion{O}{6} might be a useful signpost for
a non negligible fraction of the missing baryons, the
 bulk of the WHIM cannot be investigated through 
UV spectroscopy. X-ray band observations are required.
 
In the soft X-ray band, the residual unresolved background after 
point-sources subtraction possibly constitutes the second observational 
evidence of the WHIM. So far the tightest constraints have been obtained 
by \cite{2007ApJ...671.1523H,2007ApJ...661L.117H} 
after removing all contributions
associated to X-ray, optical, and IR sources in the {\small Chandra} Deep
Fields.  Their analysis has shown that the residual background in the
0.65--1 keV energy band cannot be accounted for by AGN below the
resolution limit. Instead, it appears to be consistent with the expected integrated
emission contributed by the WHIM, as predicted by
models based on hydrodynamical simulations
\citep[e.g.,][]{2006MNRAS.368...74R}. The 
intensity of the diffuse X-ray background 
constitutes a weak constraint on the WHIM properties.
Additional constraints can be obtained by identifying the WHIM signature
in the angular correlation properties of the  X-ray background, like in
the recent analysis of  
\cite{2009ApJ...695.1127G}.

Hydrodynamical simulations show that the best
way to investigate the WHIM is by observing the X-ray emission or
absorption lines of highly ionized element like C, N, O, Ne, Mg, and Fe.
Deep X-ray spectroscopy has indeed been performed to detect such
lines. 
>From a
theoretical viewpoint the possibility of detecting and studying the WHIM
in the absorption spectra of X-ray bright objects like AGN or Gamma-ray
bursts has
been thoroughly discussed both from a general perspective 
\citep[e.g.,][]{2002ApJ...571..563K,2003ApJ...596...19K,2003ApJ...594...42C,
2005MNRAS.360.1110V,2006ApJ...650..573C}
and by considering observations carried out with some
specific instrument 
\citep{2003MNRAS.341..792V,2003PASJ...55..879Y,2004PASJ...56..939Y,
2006PASJ...58..657K,2008SSRv..134..405P,2009ApJ...697..328B}.
>From the observational side, the analysis of the absorption spectra has lead to 
several claimed WHIM 
detections
\citep[e.g.,][]{2002ApJ...565...86F,2003ApJ...582...82M,2005Natur.433..495N}
whose statistical significance has been
questioned by subsequent studies 
\citep{2007ApJ...656..129R,2006ApJ...652..189K},
reflecting the fact that these measurements are at the limits of
current instrumental capabilities.  
The attempts to detect absorption
lines associated to dense parts of large-scale structure leads to
marginal detections with less than 4$\sigma$ significance 
\citep{2004PASJ...56L..29F,2007ApJ...655..831T,2009ApJ...695.1351B,2010ApJ...714.1715F, 2010ApJ...717...74Z}.

Studying the WHIM in absorption offers advantages and disadvantages. 
The advantage is that line
detection depends on the source flux with no need to worry about
foreground contributions. In addition, the line strength depends on the
density of the intervening absorber, allowing one to probe regions of
moderate overdensity.
The obvious drawback is that one can only probe the WHIM along the few lines
of sight to X-ray bright objects, making it very difficult to trace its
spatial distribution 
\citep[but see][for a possible way of probing
 the WHIM clustering in absorption]{2003MNRAS.341..792V}.

On the other hand, studying the  WHIM in emission is considerably 
harder but potentially more
rewarding. 
The main difficulty is represented by the fact that line detection
depends now 
 on the capability of modeling the
various contributions to the soft X-ray foreground, 
which, unfortunately, are poorly constrained by observational data 
\citep{2002ApJ...576..188M,2005ApJ...623..225S,2007ApJ...661..304H,2007PASJ...59S.141S,2008ApJ...676..335H,2009ApJ...707..644G,2009PASJ...61..805Y}.
It is therefore not surprising if only a few WHIM detections have been claimed
in continuum emission 
\citep{2005MNRAS.357..929Z,2007A&A...468..807M,2007ApJ...655..831T,
2008A&A...482L..29W}.
The main reward is constituted by the possibility of tracing the 3D distribution of the WHIM
\citep{2006ApJ...650..573C} and quantifying
its spatial and angular correlation properties
\citep{2006ApJ...652.1085U,2010arXiv1007.3274U}.
In addition,
since X-ray emissivity scales with the square of the gas density, we
expect the emission signal to be dominated by high density regions,
effectively complementing the absorption studies.

In this work we will investigate the possibility of detecting, studying
and characterizing the WHIM in emission.  In order to observe the weak
WHIM emission features a detector with a large grasp (effective area
$\times$ field of view) is mandatory.  In addition, to separate the WHIM
lines from other spectral features contributed by other sources a good
energy resolution is also required. A number of recently-proposed X-ray
satellite missions such as
{\it DIOS} \citep{2010SPIE.7732E..54O}, 
{\it EDGE} \citep{2009ExA....23...67P}, 
{\it Xenia} \citep{2010SPIE.7732E..55B}, and
{\it ORIGIN}
meet these requirements.
We will
assume observations with the detectors on board the proposed
{\it EDGE} and {\it Xenia} missions. The detailed characteristics
of these two X-ray satellites are presented in 
\cite{2009ExA....23...67P}\footnote{Also see \url{http://projects.iasf-roma.inaf.it/EDGE}},
and in \cite{2010SPIE.7732E..55B}.
For our purposes, the most relevant instrument is the
cryogenic microcalorimeter-based spectrometer array: CRIS
(Cryogenic Imaging Spectrometer).
CRIS has the energy range of
0.2--2.2~keV (goal 0.1--3~keV), the energy resolution of
$\Delta E$=2.5~eV (goal 1~eV) at 5.9~keV,
the FOV of $0.9^{\circ} \times 0.9^{\circ}$
(goal $1.1^{\circ} \times 1.1^{\circ}$),
the effective area $A=1000$ cm$^2$  (goal 1300 cm$^2$) at 0.5 keV,
and the angular resolution (HPD) of $\Delta \theta =4'$
(goal $2.5'$).
{\it DIOS} spectrometer has a similar energy resolution but
a factor of 10 smaller grasp than {\it CRIS} onboard
{\it EDGE} or {\it Xenia}.
{\it ORIGIN} is planned to have a better angular resolution
$<30''$, at the expense of energy resolution ($\Delta E$=3--5 eV).
In this work we refer to the adopted baseline of the {\it EDGE} and 
{\it Xenia}
spectrometer as a reference case and explore the effect
of a different
energy resolution, chosen in the range [1-7] eV.
In contrast,
we do not explore the angular resolution better than $1.3'$,
because the typical angular size of the WHIM above detection
threshold is $>2'$--$3'$.

The layout of this paper is as follows.  In \S~\ref{sec:whimodel}
we introduce the WHIM model used in this work from which we
obtain the simulated surface brightness  maps and emission spectra
analyzed in this work.
In \S~\ref{sec:detection}
the
problem of detecting the WHIM emission lines and computing
their relevant
statistics is addressed.  In \S~\ref{sec:phys}
we study the physical properties of the
regions in which the WHIM can be detected in emission and assess the
possibility of measuring the gas temperature from line ratios.
In \S~\ref{sec:3d} 
we investigate the possibility of tracing the spatial distribution of the
WHIM.
In \S~\ref{SEC:cons-non-ideal},
we quantify the impact of energy resolution and line contamination
form different sources. 
In \S~\ref{SEC:capab-dist-diff},
we address the question of to what extent
we can discriminate among different models of stellar
feedback and metal diffusion,  using emission studies.
Finally we discuss the main results and conclude in
\S~\ref{sec:disc}.

\section{Modeling the WHIM}
\label{sec:whimodel}

Hydrodynamical simulations provide the best way of studying the
properties of the WHIM
\citep[see, e.g.,][]
{1999ApJ...519L.109C,1999ApJ...514....1C,2001ApJ...552..473D,
2003ApJ...594...42C,2004MNRAS.348.1078B,
2006ApJ...650..560C,2008MNRAS.387..577O,
2010MNRAS.407..544B,2010MNRAS.402.1911T}.
Their
analysis has revealed that the missing baryon problem finds its natural
solution in a standard $\Lambda$CDM framework in which the WHIM is
heated up to temperature of $10^5$--$10^7$ K mostly by means of hydrodynamical
shocks resulting from the build-up of cosmic structures at scales that
are entering the nonlinear regime of density fluctuations growth. 
Additional heating/cooling mechanisms
like stellar and AGN feedback, galactic superwinds, and radiative
cooling,
and ionization by background radiation
 further
modify the thermal state of the gas.  Despite the recent advances in
modeling these mechanisms 
\citep{2006ApJ...650..560C,2006ApJ...650..573C, 2010MNRAS.407..544B, 
2010MNRAS.402.1911T,2010MNRAS.407.1581S,
2009MNRAS.393...99W,2009MNRAS.395.1875O,2010arXiv1009.0261S,
2011MNRAS.tmp..165T}
we still do not have a firm physical understanding of the effects of stellar
feedback, chief among which is metal enrichment. 

Indeed, the metal enrichment is
so important in determining the observational properties of the WHIM
that, following \cite{2009ApJ...697..328B},  we prefer  to specify gas
metallicity after the simulation run (i.e in the post-processing phase)
rather than self-consistently during the numerical simulation.
The resulting WHIM model has been shown to fulfill all existing observational 
constraints ranging from the  $dN/dz$ of the \ion{O}{6} and \ion{O}{7} 
absorption lines to the surface brightness of the diffuse X-ray background.
So, although effective, the WHIM model used in this work is 
not self-consistent. However, given the existing uncertainties in predicting 
metal abundances, in this work we will not rely on a
single, sophisticated WHIM model. Instead, in an attempt to account for
theoretical uncertainties, we will consider two WHIM models that have
been obtained from the same hydrodynamical simulations but adopting
different prescriptions for the gas metallicity.

Our WHIM models have been obtained from the hydrodynamical simulation of
\cite{2004MNRAS.348.1078B}
performed with the GADGET-2 Lagrangian code
\citep{2005MNRAS.364.1105S}
in a computational box of size $192h^{-1}$ comoving Mpc,
loaded with $480^3$ Dark Matter and $480^3$ gas particles.  
Dark matter
 and gas particles have masses of $4.62\times 10^9~h^{-1}M_\odot$
 and $6.93\times 10^8 h^{-1}~M_\odot$, respectively.
The
Plummer-equivalent gravitational softening is $\epsilon$=7.5 $h^{-1}$
kpc at $z=0$, fixed in physical units between $z=2$ and $z=0$, i.e., 
the redshift interval relevant for the study of the WHIM.  The background
cosmological model is a flat $\Lambda$CDM with $\Omega_{\Lambda}=0.7$,
$\Omega_{\rm b}=0.04$, $H_0=70 \ \kms \ {\rm Mpc}^{-1}$ ($h=0.7$), and
power spectrum normalization $\sigma_8=0.8$.  This numerical experiment
uses simple recipes to account for the main non-gravitational heating
and cooling mechanisms, namely: {\it i)} the star formation process that
is treated by adopting a sub-resolution multiphase model for the
interstellar medium 
\citep{2003MNRAS.339..289S}, {\it ii)} the feedback
from SNe that includes the effect of weak galactic outflows and {\it
iii)} the radiative cooling assuming zero metallicity and
heating/cooling from the photo-ionising UV background by 
\cite{1996ApJ...461...20H}. Finally, to construct our mock emission spectra
we have considered all available simulation outputs out to $z = 0.5$.

As mentioned above, 
the gas metallicity has been specified in the post-processing phase
and, to account for the scatter in model predictions, 
we have explored two different scenarios.
In the first one, which we call model B1, we have adopted the
deterministic
metallicity-density relation 
$Z=\min \ (0.3,0.005(1+\deltag)^{1/2})$, 
where $\deltag$ is
$\rhogas/\langle\rhogas\rangle-1$, that 
\cite{2001ApJ...557...67C} 
have proposed to match the average 
relation measured in numerical experiments.  
This is a conservative model that typically underestimates the 
line emissivity since it ignores the large scatter in the $Z-\deltag$
relation which would enhance the contribution of the line emission from 
regions characterized by moderate gas overdensity, like the WHIM.
For this reason, the B1 model should be regarded as a generous lower bound
to our detectability estimates.
In the second model, dubbed B2, we do account for the large scatter in the 
metallicity-density relation. This scatter reflects the fact that the metal
abundance at a given location is not  simply determined by the underlying
gas density but also by the thermal and chemical history of the
gas particles. Since it would be hard to model this scatter analytically
we resort to hydrodynamical simulations. For this reason we have 
enforced the same $Z-\deltag$ relation measured by 
\cite{1999ApJ...519L.109C,1999ApJ...514....1C}
in their numerical experiment.  In practice we
use the density-metallicity scatter plot measured at $z=0$ in the 
\cite{1999ApJ...514....1C}
simulation to derive a 2D probability distribution
which is then fed into a Monte Carlo procedure to assign metallicity to
gas particles with known density in the 
\cite{2004MNRAS.348.1078B}
simulation.  The result is a
metallicity model with the same $Z-\deltag$ relation
plotted in Fig. 2 of 
\cite{1999ApJ...519L.109C}
and characterized by a 1 $\sigma$ scatter of $\sim0.3$ dex.
Note that we did not use the metallicity of the \cite{2004MNRAS.348.1078B} simulation
itself since
the metal diffusion mechanism built-in the SPH code reproduces the
correct metal content in high density regions, like the intracluster
medium, but systematically underestimates the metal abundance in the
typical WHIM environments.

We stress the fact that the metallicity assigned in 
the post-processing is not self-consistent with the thermal state 
of the gas which, in the simulation considered in this work, 
is determined ignoring the role of metal cooling. In fact, recent 
numerical experiments show that the thermal state of the diffuse 
gas can be significantly altered by metal cooling
\citep{2009MNRAS.393...99W,2009MNRAS.395.1875O,2010arXiv1009.0261S,
2011MNRAS.tmp..165T}.
However, including the effect of metal cooling is beyond the scope of this paper.
Instead, we attempt to bracket the uncertainties that derives from our simplified 
by adopting the two very different density-metallicity relations, being confident that
the deterministic density-metallicity relation will provide very conservative 
predictions for the detectability of the WHIM.

To determine the ionization balance we have assumed pure collisional ionization 
equilibrium (CIE) and not  hybrid collisional and
photoionization equilibrium as in \cite{2009ApJ...697..328B}.
The use of the  CIE assumption is justified by the fact that in this work
we are interested in emission spectra
whose intensity 
scales with the squared
density of the gas and thus it is heavily weighted towards high density
peaks where the ionisation is basically determined by collisions.
\cite{2009ApJ...697..328B}  were interested in absorption lines 
which typically trace
the WHIM in regions of moderate overdensity 
where  photo-ionization cannot be neglected. Here
we focus on  WHIM lines 
For this reason we have assumed CIE 
and used the APEC thermal
model included in the code XSPEC 
\citep{2001ApJ...556L..91S}
 to compute emission spectra \citep{2006ApJ...652.1085U}.
The resulting mock X-ray spectra contains all metal lines but here 
we focus on the strongest ones only: the 
\ion{O}{8} Ly$\alpha$ line ($E=$0.653 keV) and 
the \ion{O}{7} K$\alpha$ triplet (E= 0.561, 0.569, and 0.574~keV).

Model B2 agrees 
with all available observational constraints
and, from a theoretical viewpoint, predicts a  phase-space diagram 
(shown in Figure~6 of  \cite{2009ApJ...697..328B}) which is 
remarkably close to that of  
\cite{2006ApJ...650..560C} (shown 
in Figure~6 of that paper),
a fact that further increases our confidence on  model B2.
Therefore in this work we consider B2 as the reference model and
regard predictions from model B1 as a generous lower bound.

One final caveat for model B2: this model is known to overestimate the
X-ray emissivity in high ($\deltag>10^3$) density environments, i.e.
emission from gas in groups and clusters \citep{ACFpaper}.  While the
precise reason for this systematics is unknown, a plausible explanation
is represented by the fact that in our model we have assumed primordial
composition in the cooling function, instead of updating the cooling
rate according to the evolving metallicity of the gas.  The analysis of
{\cite{2010MNRAS.407..544B} suggests that this would artificially
increase the emissivity in regions with high density.
Brighter emission from group-like environments increases the chance
superposition from strong emission lines, like those of the Fe L
complex, that may contaminate or even outshine the weaker O lines,
artificially reducing our theoretical estimate of WHIM detection.
We will quantify these effects in \S~\ref{SEC:detect-emiss-source}.
We again stress the fact that the inclusion of metal line cooling 
might have an impact on both the line surface brightness and the thermal state 
of the gas. We qualitatively discuss the expected impact of metal cooling  
in \S~\ref{sec:disc}.

\subsection{Mock emission  spectra}

\label{sec:mockspectra}

\begin{figure}
\epsscale{.98}
\plotone{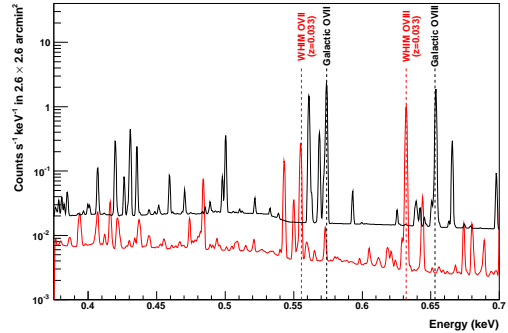}
\caption{
Emission spectrum  in a 
$2.6'\times2.6'$
area 
taken in a 1 Ms observation with {\it Xenia} CRIS, assuming an energy resolution 
$\Delta E=1~\mathrm{eV}$. 
Black: Sum of the Galactic Foreground and unresolved
extragalactic background. The Galactic \ion{O}{7} triplet
and the \ion{O}{8} K$\alpha$ line at $z=0$ are flagged.
Red: Contribution from the extragalactic gas.
The  \ion{O}{7} and \ion{O}{8} lines
at  $z=0.033$ are indicated in the plot.
\label{fig:emspec}
}
\end{figure}

The detailed procedure used to produce the mock spectra is described in detail 
in  \cite{2010arXiv1007.3274U}, here we provide a brief summary.
To compute the mock X-ray spectra we have considered all gas particles 
within a light-cone extracted from the hydrodynamical simulation, out to 
$z=0.51$. The light cone is formed by stacking the simulation outputs corresponding
to 7 redshift intervals, each one of them corresponding to a comoving depth
of 192$h^{-1}$~Mpc. 
We do not consider the WHIM at  $z > 0.5$ since in that case 
 the \ion{O}{7} lines would be redshifted to energies at which 
a number of emission lines due to L and M transitions of heavier
metals make identification of O lines difficult.  
In the stacking process, the simulation cubes that constitutes the 
individual redshift intervals were
randomly shifted and rotated to avoid periodic replicas of the same
large scale structures.  
Then, after selecting a random observer within
the cube at $z=0$ we generate a  single cone of view of $5.5^{\circ} \times
5.5^{\circ}$ and tag gas particles with their redshift, angular position, 
density, and temperature.  
 Particles with temperature in the range $10^5-10^7$ K
and overdensity $\delta<1000$ are classified as ``WHIM phase''.  
Among the remaining gas
particles, those  
that provide a significant contribution to line emission are
either in a ``hot phase'' ($\delta \ge 1000$ and $T \ge 10^7$ K, roughly
corresponding to gas within clusters of galaxies) or in a ``dense phase''
($\delta \ge 1000$, $T < 10^7$ K, typically associated to galaxy
groups).  Their contribution to the X-ray emission is included 
in our mock spectra. The contribution from particles 
colder than  $10^5$ K is negligible.
In the mock spectra we did not include the  contribution from gas at $z>0.51$. 
Emission lines (like the Fe L complex in galaxy groups) from high-$z$ gas  
potentially contaminates spectra in the energy range in which we search for
WHIM lines. We estimate the impact of contamination by spurious lines
in \S~\ref{SEC:detect-emiss-source}.

To account for the finite angular resolution of the instrument, the
 FOV of the light cone is 
divided into $256 \times 256$ pixels of $1.3' \times 1.3'$,
i.e. smaller than the planned CRIS resolution but possibly 
consistent with alternative instrument designs. 
The mock X-ray
spectrum within each angular resolution element was obtained by 
adding together the individual spectra from each gas particle
within the pixel plus that of gas particles in the contiguous pixels, weighted
according to the smoothing kernel.
The mock spectra were then sampled with an energy resolution of 1 eV
to mimic the energy resolution of the instrument.
The effect of the Galactic absorption has been
included using the model of 
\cite{1983ApJ...270..119M}
in which we have
adopted typical high latitude column density value of $1.8 \times
10^{20}~\mathrm{cm}^{-2}$ 
\citep{2002ApJ...576..188M}.
We have assumed the
same metal solar abundance as in the 
\cite{1989GeCoA..53..197A} models.
The final result is a suite of $256 \times 256\times7$ mock spectra
that we search for detectable WHIM lines.
A typical example of mock spectrum 
is shown in Fig.~\ref{fig:emspec}.
The red curve represents the contribution by extragalactic gas in a
light cone with FOV of $2.6'\times2.6'$ (i.e. the sum 
of $2 \times 2 \times7$ individual mock spectra). Photon counts
have been computed in $\Delta E= 1$ eV energy bins assuming a 1 Ms
observation with CRIS. The spectrum is convolved with the detector
response matrix of 1~eV energy resolution (FWHM).  
Emission lines from different metals are clearly
visible. In particular we have flagged those lines corresponding to a
line system at $z=0.033$. The \ion{O}{7} triplet and the \ion{O}{8}
recombination
line are clearly visible.  The black curve shows the sum of the
contributions of the Galactic foreground and the unresolved
extragalactic background modeled after 
\cite{2002ApJ...576..188M}.  The
Galactic \ion{O}{7} and \ion{O}{8}
 lines at $z=0$ are clearly seen in correspondence
of their rest frame energy.

\section{Analysis of the Mock Spectra}
\label{sec:detection}

To assess the possibility of detecting the WHIM in emission 
and constraining its thermal state with next-generation instrument
we consider spectra in larger pixels ($2.6'\times2.6'$) 
similar to the CRIS angular resolution (goal).
For this purpose we search for emission lines in each of the 
 $128 \times128 \times 7$ mock spectra using a two step procedure.
{\it (1)} We identify the \ion{O}{7} (triplet) and \ion{O}{8}
lines by searching for local maxima 
in the appropriate energy
range of the spectrum.  
{\it (2)} we compute the surface brightness of each line by summing
the flux over energy bins redward and blueward of the line maximum
and stop whenever the surface brightness SB$_{th}$ in the bin drops 
below $7\times10^{-3}$~\phot\ 
which is smaller than the typical line detection threshold 
for a 1 Ms observation with CRIS, as we will show 
\S~\ref{sec:linedet}.
A simple example serves to clarify the procedure. To identify
\ion{O}{8} lines produced in the redshift range $z=[0,0.065]$ 
we search for local maxima in the energy range [0.613, 0.653] eV, where 
the lower bound is the redshifted energy of an \ion{O}{8}
 photon emitted at $z=0.065$
The centroid of the line is identified with the maximum and its intensity 
 is obtained by summing over all contiguous energy 
bins with surface brightness above that of the selected threshold.
This strategy,  which minimizes the chance contamination from
gas at different redshifts, cannot be applied to real spectra since we
lack the information of
the redshift of the gas responsible for the line emission. 
This effect is discussed
in \S~\ref{SEC:detect-emiss-source}.

Detectable WHIM lines are rare and it is unlikely to find two of them
in the same mock spectrum. On the other hand, when approaching the 
surface brightness threshold, several weak  emission lines start to appear.
They are typically clustered around stronger lines and thus are 
likely to be physically associated to the same line emitting element.
Considering all these lines would spoil line statistics since they 
would oversample the same line-emitting regions.
To circumvent this problem we have adopted a simple
criterion: we merge together all lines that are separated 
by a number of energy bins smaller than some minimum
amount $\nbin$.
In our analysis we have adopted $\nbin=8$ (corresponding  8~eV)
after having checked that the number of  lines with
0.1~$\mathrm{ph~s^{-1}~cm^{-2}~sr^{-1}}$ (strong enough to be hopefully
detected by next-generation 
instruments) depends neither on  $n_{\rm bin}$ nor on  SB$_{th}$.

To confirm that our B2 model fulfill the current constraints on the 
soft X-ray diffuse background, we have compared the total surface brightness
predicted by the model in the  [0.65, 1] keV band with that measured by
\cite{2007ApJ...671.1523H} after removing the contribution from all known sources.
According to the model, the WHIM contribution to the surface brightness contributed 
is $(3.6\pm 0.3) \times 10^{-13}$ erg cm$^{-2}$ s$^{-1}$
deg$^{-1}$, safely below the observed value of $(1.0 \pm 0.2) \times
10^{-12}$
erg cm$^{-2}$ s$^{-1}$ deg$^{-1}$. 
The more conservative B1 model, which predicts a lower surface brightness 
is also obviously in agreement with observations.

\subsection{Line Statistics}
\label{sec:linestat}

\begin{deluxetable*}{ccccccc}
\tablecaption{Emission line detections
\label{tab:tab2}
}  
\tablewidth{0pt}
\tablehead{
\colhead{Model} &
\colhead{$t_{\rm exp}$} &
\colhead{$f_{\rm line}$} &
\colhead{$dN_{\rm OVII}/dz$} &
\colhead{$dN_{\rm OVIII}/dz$} &
\colhead{$dN_{\rm OVII+OVIII}/dz$} &
\colhead{$N_{\rm OVII+OVIII}$  per deg$^2$}
}
\startdata
B2 & 1.0 & 0.07 & 4.2 (3.1) & 2.6 (1.9) & 2.4 (1.6) & 639 (426) \\
B2 & 0.1 & 0.48 & 2.1 (0.6) & 1.5 (0.5) & 1.1 (0.2) & 293 (53) \\
B1 & 1.0 & 0.07 & 3.3 (1.6) & 2.1 (1.0) & 1.8 (0.7) & 479 (186)\\
B1 & 0.1 & 0.48 & 1.4 (0.03) & 1.1 (0.05) & 0.7 ($<10^{-3}$) 
 & 186 (0)\\
\enddata
\tablecomments{
Column 1: WHIM  model.
Column 2: Exposure time (Ms).
Column 3: Minimum line surface brightness required for a 5$\sigma$ detection
(\phot).
Column 4: Expected number of \ion{O}{7} detections per resolution element
and unit redshift 
Column 5: Expected number of \ion{O}{8} detections per resolution element
and unit redshift. 
Column 6: Expected number of simultaneous  \ion{O}{8} and \ion{O}{8} detections 
per resolution element and unit redshift. 
Column 7: Expected number of simultaneous \ion{O}{8} and \ion{O}{8} detections 
per square degree due to gas within $z=0.5$.
All estimates assume an angular resolution of $2.6'\times 2.6'$.
The numbers in parenthesis indicate lines contributed by the WHIM.
}
\end{deluxetable*}

\begin{figure}
\centering
\includegraphics[height=0.95\columnwidth,angle=-90]{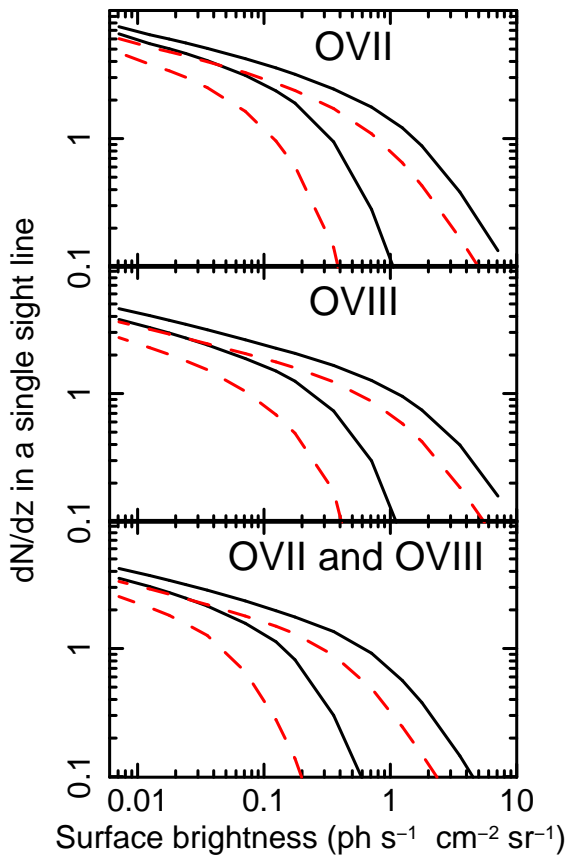}
\caption{Cumulative number of emission lines per unit redshift as a function of line 
surface brightness. 
Each panel refers to
\ion{O}{7} triplet (top), \ion{O}{8} Ly$\alpha$ (middle) or
both \ion{O}{7} triplet and \ion{O}{8} Ly$\alpha$ (bottom).
Black, solid curves indicate model B2, while red, dashed curves
model B1. 
For each model, upper curve shows all the gas and 
lower curve refers to ``WHIM phase'' gas only.
\label{fig:dndzo7o8o78}}
\end{figure}

After having identified all \ion{O}{7} and \ion{O}{8} emission lines in the mock
spectra we can compute their cumulative number per unit redshift as a
function of the line surface brightness.  The top panel of
Figure~\ref{fig:dndzo7o8o78} shows
the case of the \ion{O}{7} triplet whose surface brightness is the sum of the three lines. 
The black continuous
curves refer to model B2 while the red dashed curves refer to model B1. For
each model, the upper curve accounts for the gas emission from all gas
particles along the line of sight and the lower curve considers only the
contribution from the
``WHIM phase'' (see \S~\ref{sec:mockspectra}).
In the B2 model the lines contributed by the warm hot gas
constitutes a significant fraction of the total. However, this fraction
decreases with the surface brightness of the lines. It 
ranges from $\sim 100$ \% for the weakest line to $\sim
5\%$ for the brightest one.  This result shows that the possibility 
of detecting the WHIM emission lines increases when decreasing the line 
detection threshold and  clearly illustrates the need of increase the sensitivity of 
next-generation instruments.
Model B1 predicts $\sim 30 \%$ less lines than
model B2, as expected.  The difference among the models 
is almost independent of the line surface brightness.
For a reference line surface brightness of 
0.1~\phot\
(matching the  typical detection threshold for next-generation
X-ray spectrometers as we will show) 
the expected number of detected lines 
is of order of unity, indicating  that the chance of multiple line
detections in a pixel is rather small.

The middle panel of 
Figure~\ref{fig:dndzo7o8o78} shows the  cumulative $dN/dz$ statistics
for the \ion{O}{8} line. All considerations made for the \ion{O}{7} lines apply to 
the \ion{O}{8} case except for the fact that, for a given surface brightness 
the expected  number of \ion{O}{7} triplets is larger than that of  \ion{O}{8} lines.
This difference  becomes smaller if one considers  
the brightest \ion{O}{7} lines rather than the triplet.
The bottom panel shows the cumulative $dN/dz$ of those 
line systems that exhibit both the \ion{O}{7} triplet and \ion{O}{8} line
as a function of the total surface brightness.
The curves are similar to those of the \ion{O}{8} lines (middle panel)
except that at the bright end. This is due to the fact that 
presence of an \ion{O}{8} line  almost invariably guarantees that of an
\ion{O}{7} triplet with similar (or larger) 
surface brightness.

Our results can be compared with the theoretical predictions of 
\cite{2006ApJ...650..573C}
who used more sophisticated models that account for
departures from ionization equilibrium and the effects of galactic
superwinds.  
Their predicted cumulative number of emission lines (see Fig.~12 of the paper) 
is similar to that predicted by model B2.
In particular the agreement is remarkable for the \ion{O}{8}  lines, while 
we seem to over predict that of \ion{O}{7} triplets.
In addition, we note that the difference between model B1 and B2 is smaller
than the scatter in their model predictions, indicating that we are being generous 
in our attempt of bracketing theoretical uncertainties.

\subsection{Expected WHIM detections}
\label{sec:linedet}

To detect  \ion{O}{7} and \ion{O}{8} emission lines one needs to discriminate
the genuine, weak emission signal from the astrophysical foreground and
background which represents the main source of noise,
typically exceeding the instrumental background.  The problem of
detecting the WHIM emission lines is analogous to that of detecting a low-surface
brightness source with a typical angular extension is a few arcmin, 
i.e., the typical angular size of a WHIM filament at $z\sim 0.3$,
The signal to noise is then optimized by measuring spectra with an angular resolution
comparable to that of the sources over large FOV to detect as many WHIM lines 
as possible. In practice, one wants to perform spatially resolved spectroscopy 
over large areas with an angular resolution of few arcmin.
The CRIS detector proposed for the {\it Xenia} mission and that we are considering as 
a prototype for next-generation spectrographs
fulfills these requirements.

To investigate the possibility of detecting the WHIM through line emissions
and assess the statistical significance of the detections
one needs to quantify the noise which is mainly contributed by two sources:
the cosmic X-ray background  (CXB, hereafter)
produced by unresolved extragalactic point sources,
and the foreground emission of our Galaxy.

The CXB is contributed by all X-ray sources 
(mostly AGN) below the source detection threshold.
Its intensity depends therefore on the
characteristics of the detector and on the exposure time.
A conservative upper limit to the CXB can be obtained 
from the XQC rocket experiment 
\citep{2002ApJ...576..188M} since this estimate includes 
the contribution of all point sources 
in the FOV of $\sim$1~sr.
In this case the
CXB surface brightness is 
$f_{\rm CXB} \sim 30$ \photkeV.
This value can be reduced by removing all resolved 
sources. One can remove sources previously detected in 
some other surveys. However, efficient removal would be possible only
in areas covered by deep surveys, which are typically small.
The alternative is to perform simultaneous
deep imaging and resolved spectroscopy of the same areas, a strategy
that would be possible if both a CCD-like imager and a 2D spectrograph 
with similar FOVs were available. This is precisely the case of
 of the {\it Xenia} mission concept which should be capable of 
resolving and removing $\sim1000$ 
AGN per square degree 
down to a limiting flux of
 $2.5 \times 10^{-16}$~erg~s$^{-1}$~cm$^{-2}$ in the
$0.5-2$~keV band. 
If AGNs were isotropically distributed, then a spectrograph with $\sim4'$
angular resolution would have all its pixels contaminated by AGN
emission.  Taking into account the angular clustering of the AGNs
\citep{1995ApJ...455L.109V}, this contamination would affect $\sim$63\%
of the pixels if the angular resolution of the instrument is improved to
$2.5'$ (i.e., the CRIS goal).  It is clear that the best compromise
between low AGN contamination and large number of available
pixels depends on the actual characteristics of the detector and 
on the availability of a CCD imager.
For the sake of simplicity in this work we adopt the conservative solution 
of keeping all pixels and thus we assume 
$f_{\rm CXB} = 30$~\photkeV.

The Galactic foreground is basically contributed by three sources: the
so-called local bubble, the Galactic halo, and the solar wind charge
exchange (SWCX).  Following \cite{2002ApJ...576..188M} we describe the
foreground emission as the sum of absorbed thermal model (\emph{wabs}
and \emph{apec} model in the XSPEC package respectively) and unabsorbed
thermal model (\emph{apec}).  The former component represents the
Galactic halo, while the latter the sum of local bubble and SWCX.  One
may think that this model is too crude, since the SWCX contribution,
which is non thermal, is expected to outshine that of the local bubble.
Moreover, the variable nature of the SWCX and the lack of strong
observational constraints
\citep[e.g.,][]{2008ApJ...676..335H,2007PASJ...59S.133F} make it
difficult to model its contribution.  However, for the
purpose of this work, this approximation is reasonable for two reasons.
First, it has been shown that the sum of two thermal components,
although inaccurate, is quite effective in reproducing observations, at
least within the energy resolution of CCD detectors
\citep{2007ApJ...658.1081G,2009PASJ...61..805Y}.  Second, in the
analysis of the real spectra one ignores the energy ranges in which
Galactic emission lines are present. These lines include Ly$\beta$,
Ly$\gamma$, and Ly$\delta$ transitions from H-like C and N atoms, which
should constitute strong features of the SWCX emission. In fact, all
SWCX lines listed by
\cite{2006A&A...460..289K} and \cite{2007A&A...469.1183B} with a 
contribution of $>$a few\% of \ion{O}{7} triplets are in
energy intervals that we ignore when we search for the WHIM lines in
\S~\ref{SEC:detect-emiss-source}.

\begin{figure}
\epsscale{.98}
\plotone{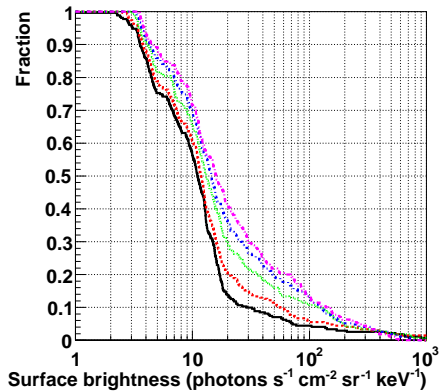}
\caption{ 
Fraction of energy bins (each bin with 1 eV width)
 in which the measured surface brightness is above the
 value indicated on the  $X$--axis.
Different curves represent different energy resolution (FWHM)
of the instrument:
1 eV (red dashed), 3 eV  (green dotted), 5 eV (blue dot-dashed),
and 7~eV (magenta dash dotted).
The black, solid curve shows the case of no convolution with the 
detector response.
\label{fig:GalacticLineSB}
}
\end{figure}

Since the energy spectrum of the Galactic foreground is characterized by
prominent emission lines (see Fig.~2) and we are interested in
identifying those produced by the WHIM, 
Galactic foreground
flux cannot be estimated by the averaged flux over the wide energy range
(e.g., 0.4--0.7 keV).  More realistically, we would need to restrict
the search to those energy bins that are not heavily contaminated by
Galactic lines.  Once again, when adopting this strategy one should
compromise between the need of keeping as many energy resolution
elements as possible and that of minimizing the Galactic signal.  The
solution can be found by looking at Figure~\ref{fig:GalacticLineSB} that
shows the fraction of energy bins in which the measured foreground
surface brightness is above the value indicated on the $X$--axis. The
shape of the 
curves that refer to the case of no convolution with the detector
response (black, solid) and of 1 eV energy resolution (red, dashed)
is characterized
by a sharp drop at
moderate surface brightness values, followed by a flattening in the high
surface brightness tail.  This bi-modality reflects the transition from
a regime in which the flux is dominated by the continuum contribution
(the low surface brightness end) to a regime dominated by the emission
of bright lines (the high surface brightness end).  The transition
occurs at a surface brightness of 
$f_{\rm FG} \sim 20$~\photkeV, 
which therefore we take as the reference
value for the Galactic foreground continuum signal in this paper.
The other curves of Fig.~\ref{fig:GalacticLineSB} refer to different
energy resolutions.  As the energy resolution degrades, the fraction
of high Galactic surface brightness becomes larger.  The influence
is considered in \S~\ref{SEC:infl-finite-energy}.


Having estimated the two noise sources
($f_{\rm CXB}$ and $f_{\rm FG}$ to 30 and 20
\photkeV, respectively) and
ignoring the negligible contribution from the instrumental noise  \citep{2007PASJ...59S..77K},
one can compute the significance of a line detection from 
the following expression:
\begin{equation}
\sigma_{\rm line}=
\frac{f_{\rm line}}{\sqrt{[f_{\rm line}+(f_{\rm CXB} +f_{\rm FG})\Delta E]}}
\sqrt{\Delta\Omega t_{\rm exp} A_{\rm eff}},
\label{eq:sn}
\end{equation}
where $f_{\rm line}$ is the source line flux in units of 
\phot,
$\Delta E$ is the energy resolution of the
instrument, $A_{\rm eff} $ its effective area, $\Delta\Omega$ its
angular size (or the pixel size in the case of this work),
 and $t_{\rm exp}$ is the exposure time.
Eq.~(\ref{eq:sn}) assumes that both signal and noise can be treated as
Poisson variables.

In Table~\ref{tab:tab2} we list for the two WHIM model explored (column
1) and two different exposure times considered in this work
(column 2), the minimum line surface
brightness required for a 5$\sigma$ line detection 
calculated with Eq.~(\ref{eq:sn})
in a
hypothetical observation performed with CRIS (col. 3).  The
corresponding number of detections per resolution element and unit
redshift can be obtained from the $dN/dz$ statistics plotted in
Fig.~\ref{fig:dndzo7o8o78}. 
They
are listed in columns 4 (\ion{O}{7}), 5 (\ion{O}{8}), and 6 (simultaneous \ion{O}{7} and
\ion{O}{8} detections).  Column 7 shows the number of
\ion{O}{7}+\ion{O}{8} WHIM detections expected in the 1 degree$^2$
 FOV due to the
gas at $z \le 0.5$.  Each entry in columns 4--6 shows two values.
The
higher one represents the expected number of detected lines contributed
by all gas, not just in the WHIM phase.  
The smaller one, in parenthesis,
refers to lines that can be attributed to the WHIM.  All detection
estimates assume an angular resolution of
$2.6'\times2.6'$ (corresponding to the sum of 2$\times$2 
pixels in the simulated dataset to the CRIS goal resolution)
and 1~eV energy resolution.
These estimates do not account for the presence of spurious emission
lines that contaminates $\sim$20\% of the energy band analyzed.  Their
impact on the ideal detection estimates listed in 
Table~\ref{tab:tab2} is estimated in \S~\ref{SEC:cons-non-ideal}.

>From Table~\ref{tab:tab2} we see that the expected number of 
detections is quite large even in the least favourable case (model B1,
$t_{\rm exp}$=100 ks, simultaneous \ion{O}{7}+\ion{O}{8} detection). 
However, only 1 Ms long observations guarantee to observe WHIM
lines in the most conservative scenario of model B1. With the more 
realistic model B2 one expects to be able to detect WHIM lines also 
with shorter (100 ks) exposure times.
We note that WHIM detection estimates in column 7 are to be regarded as 
conservative since they only consider gas out to $z=0.5$ whereas hydrodynamical
simulations show that the WHIM mass fraction is still quite large
at $z=1.0$.
Overall, these results indicate that
next-generation instruments will allow one
to unambiguously detect the WHIM in emission
and, as we will see in the next sections, to characterize its thermal state and trace its
spatial distribution.

\section{Physical properties of the WHIM in emission}
\label{sec:phys}

\begin{figure*}
\centering
\includegraphics[width=0.45\textwidth,angle=0]{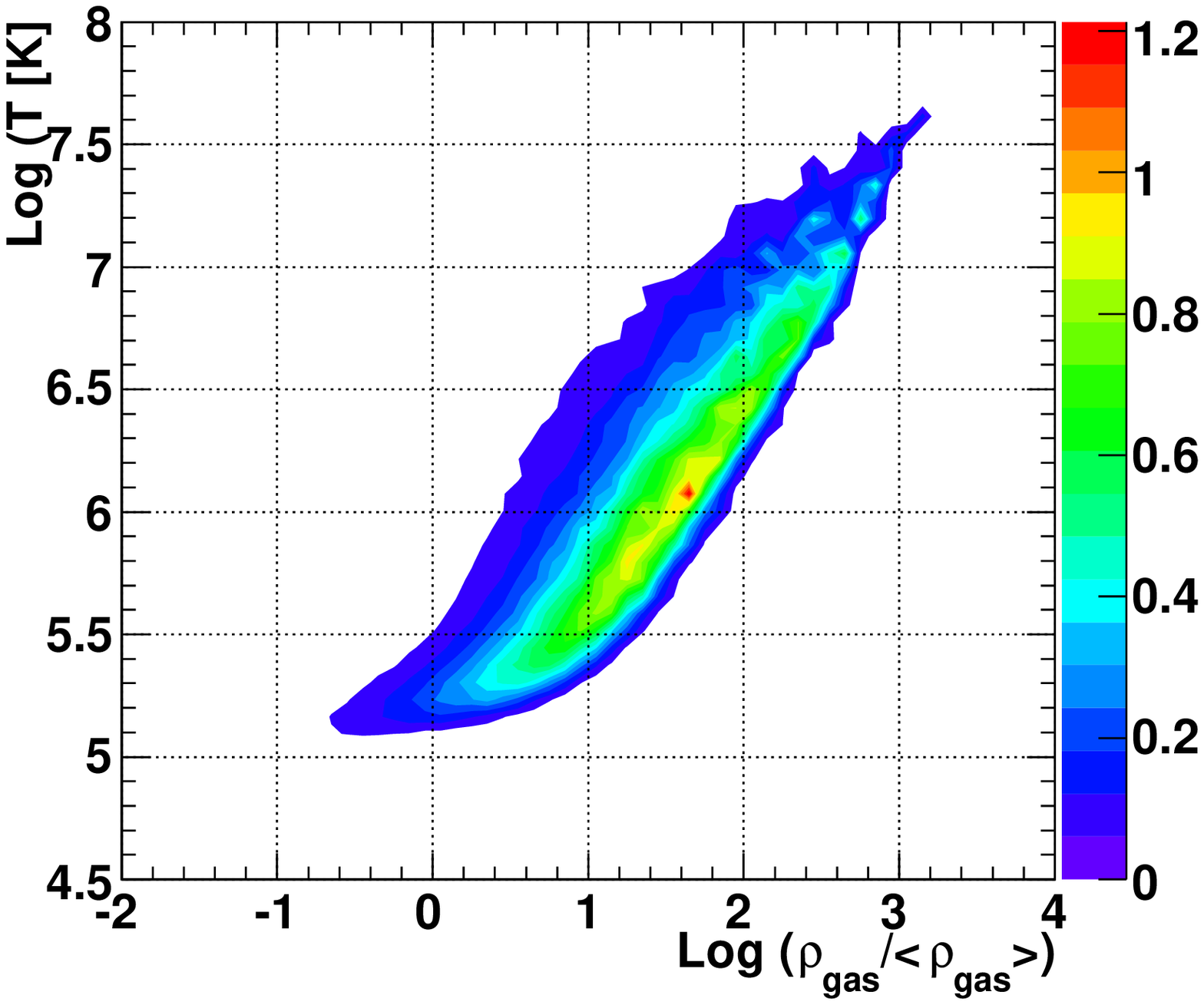}
\includegraphics[width=0.45\textwidth,angle=0]{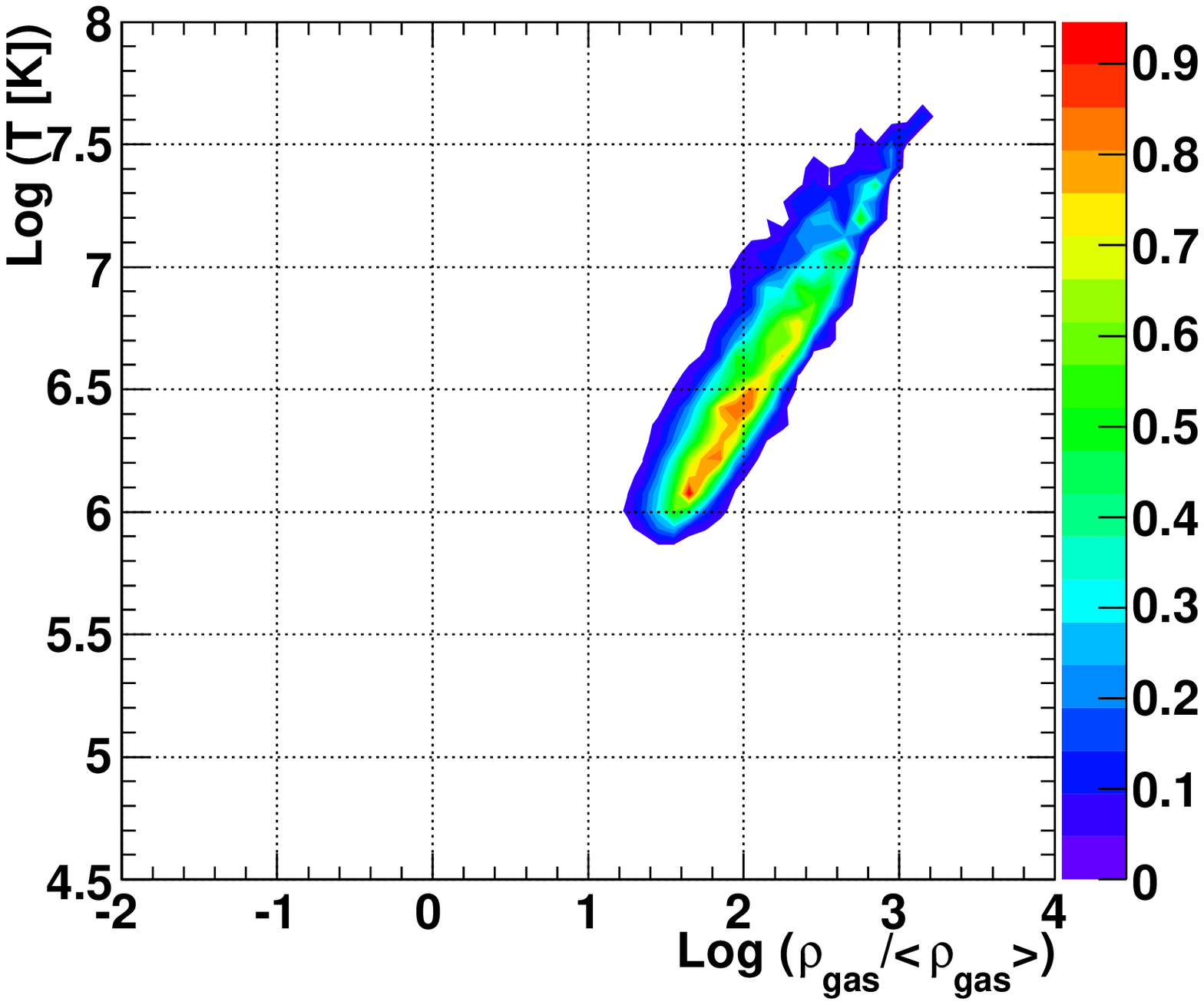}\\
\includegraphics[width=0.45\textwidth,angle=0]{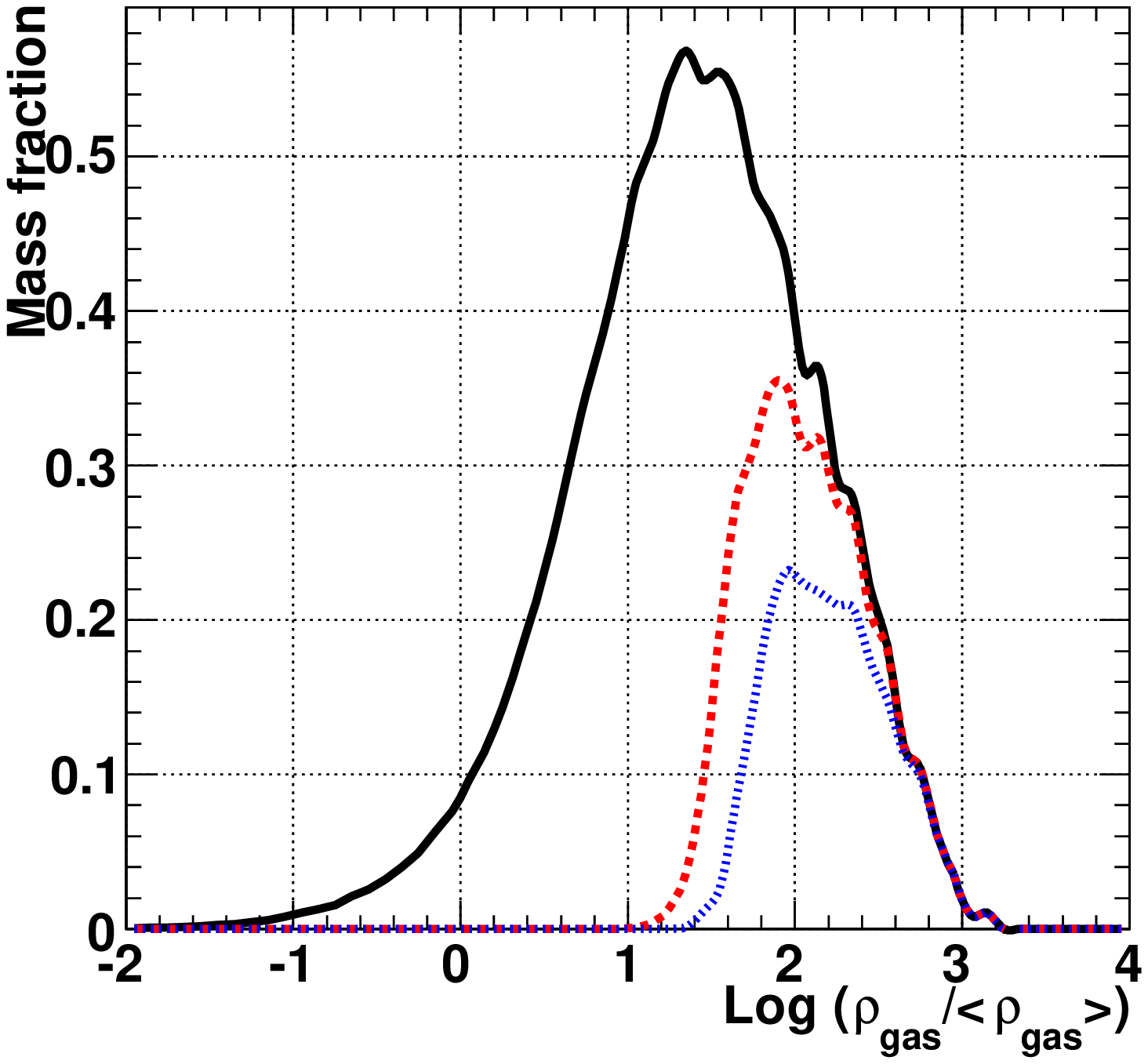}
\caption{ 
Top left: 
Contours of constant gas
mass fraction in phase-space, i.e., the fraction in
unit $\log\rho$ and $\log T$ intervals.
The gas density is
normalized to its cosmic mean ($X$--axis)
and its temperature is in unit of K ($Y$--axis).
The plots considers only gas particles in the redshift slice
$0.202<z<0.274$ and with temperature $T>10^{5}$~K.
Color coded contours are drawn in correspondence of different 
values of the gas mass fraction,  indicated in the color scale.
Top right: same as the top-left panel, but referring to gas elements 
characterized by \ion{O}{7} + \ion{O}{8} line systems strong enough 
to be detected with 1~Ms observation with CRIS.
Bottom panel: the probability distribution function of the gas density
obtained by merging the gas mass fraction in the 
$\rho$--T plane over its temperature. Black curve: all gas elements.
Red dotted curve:  gas elements in with both \ion{O}{7}
and \ion{O}{8} above 5 $\sigma$ detection threshold
of 0.07~\phot,
corresponding to 1~Ms exposure with CRIS.
Blue dashed curve:  same as the red dotted curve but 
referring to a 100~ks exposure with detection threshold of
0.48~\phot.
\label{fig:rhoTdetected}
}
\end{figure*}

\begin{figure*}
\epsscale{.95}
\plottwo{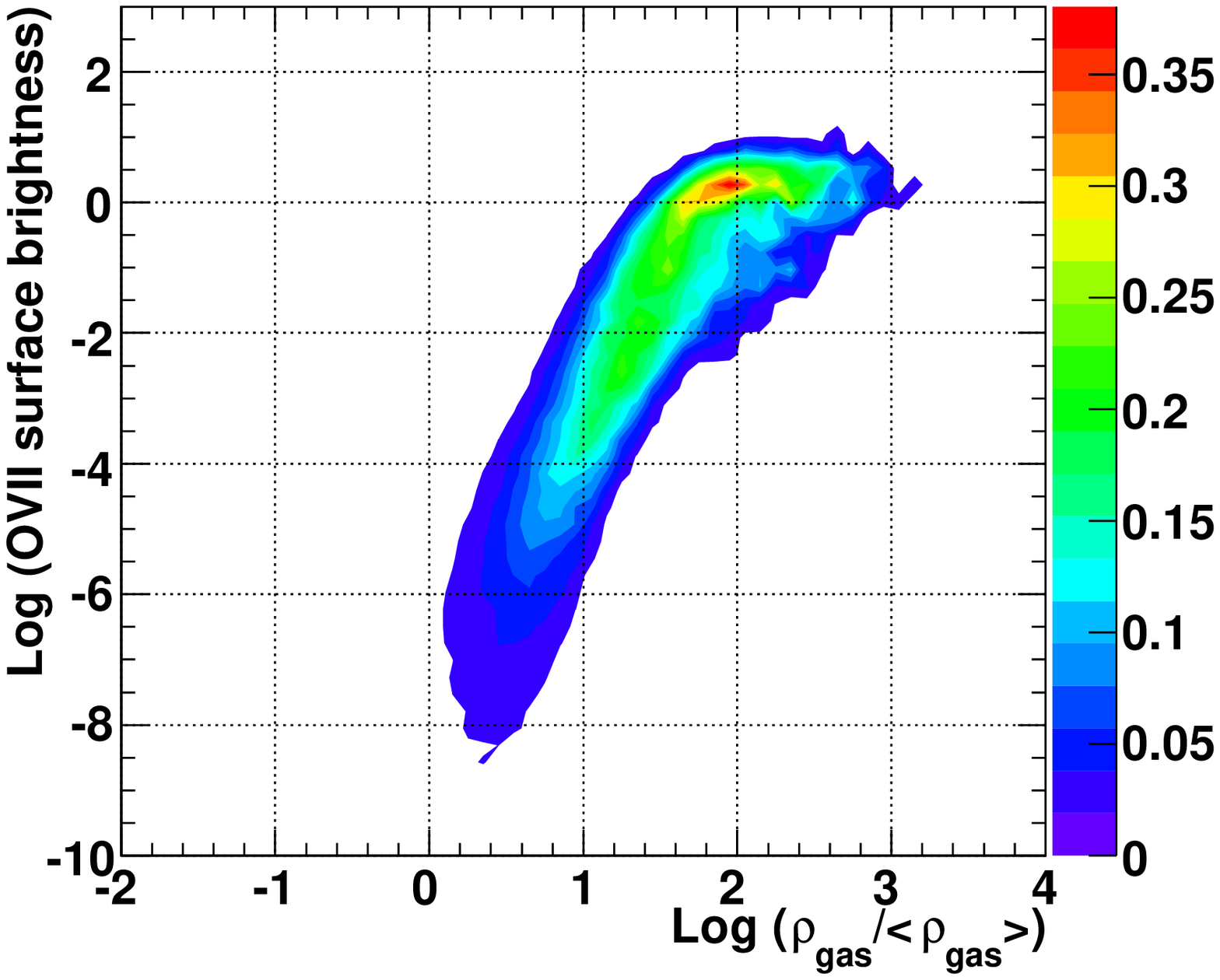}{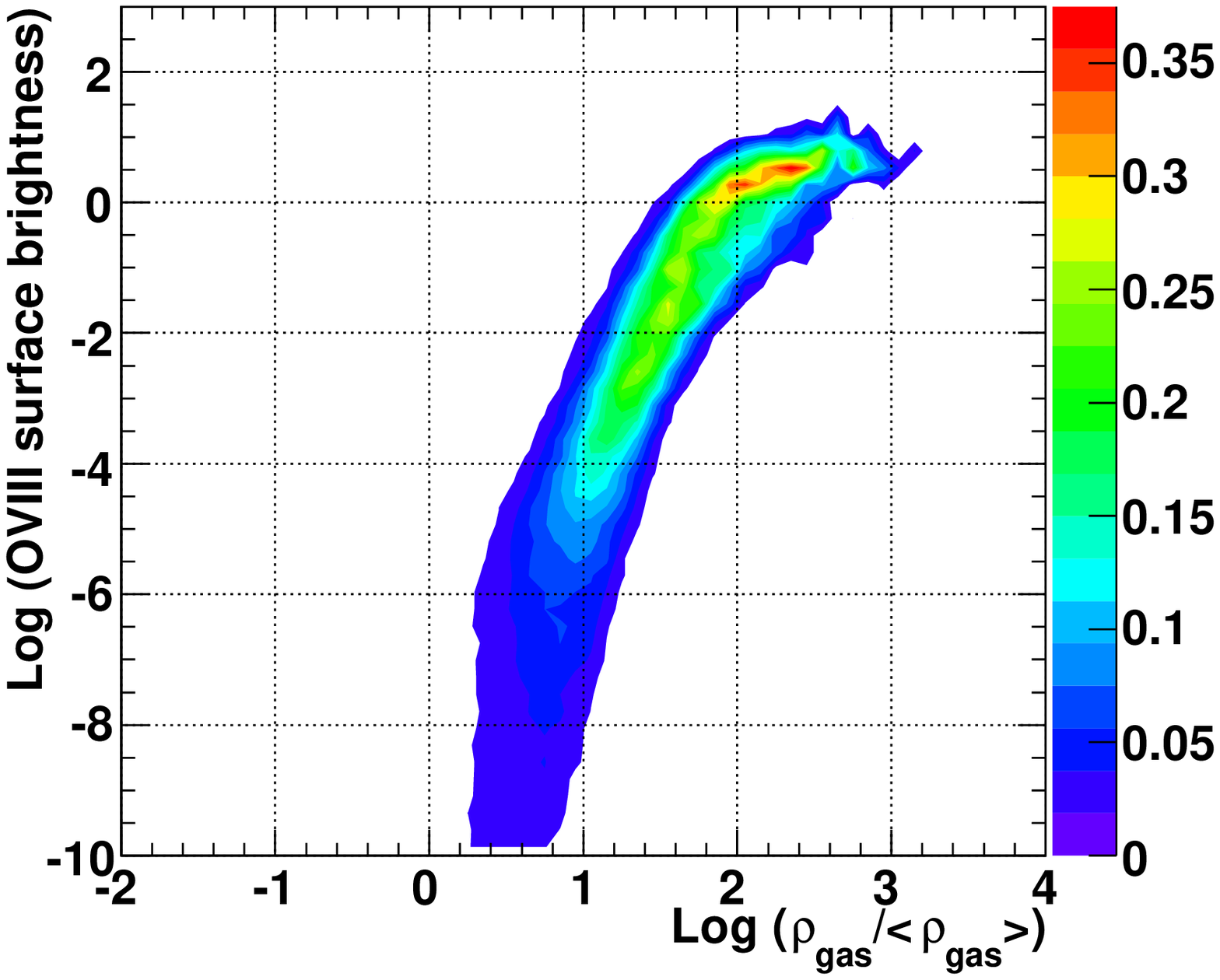}
\caption{ 
Contours of constant gas mass fraction 
in the line surface brightnesses-gas density plane. 
The plots considers only gas particles in the redshift slice
$0.202<z<0.274$ and with temperature $T>10^{5}$~K.
The surface brightness ($Y$--axis) is in units of
\phot\ and the gas density 
($X$--axis) is normalized to its cosmic mean. Left panel: 
Surface brightness of the \ion{O}{7} triplet. 
Right panel : \ion{O}{8} recombination line. 
The detection limit with a {\it Xenia} 1~Ms observation is
$0.07$~\phot,
or $Y=-1.15$.
\label{fig:SBrho}
}
\end{figure*}

In this section we investigate the physical properties (density
$\rhogas$, temperature $T$, and mass $M$) of the WHIM that,
according to our results, can potentially be detected by deep 
resolved spectroscopy in next-generation X-ray missions.
The relevant quantities obtained from
the analysis of the  mock X-ray spectra 
will be 
compared with the corresponding 
``true (input)''
physical properties of the WHIM 
measured directly from the simulation.
To perform  realistic
comparisons the ``true''
quantities were measured by averaging over all gas particles 
that contribute to the spectral signal, i.e.  particles within 
volume elements.
The data used for comparison are taken from
the FOV of $5.5^\circ\times5.5^\circ$ and the redshift range of
$0.202<z<0.274$.
The angular size of each volume element is 
$2.6'\times 2.6'$ (the angular resolution)
in \S~\ref{SEC:dens-temp-detect} and
$1.3'\times 1.3'$ in \S~\ref{sec:tgas}.
The depth of a volume element 
is 3 Mpc (comparable to the energy resolution).

\subsection{Density and temperature of  detectable WHIM}
\label{SEC:dens-temp-detect}
The thermal state of the gas can be appreciated from the phase-space
diagram shown in top-left panel of Fig.~\ref{fig:rhoTdetected}.
Colour contours are drawn in correspondence of the same gas mass
fraction, indicated in the color-scale.  This plot is analogous to the
one shown in Fig. 6 of 
\cite{2009ApJ...697..328B} with some important
differences.  First of all here we only
consider  particles with $T>10^{5}$~K
that are potentially relevant for the line emission.  Second, this plot
refers to redshift slice $0.202<z<0.274$, considerably narrower 
than $0<z<0.2$, the one considered by \cite{2009ApJ...697..328B}.
Finally, and most important, the plot refers to gas
density and temperature averaged in the 
$2.6'\times 2.6'\times3$ Mpc volumes and not to 
the density and temperature of each gas particle
in the hydrodynamical simulation, which cannot be
observed directly.
The simulated data considered in Fig.~\ref{fig:rhoTdetected}
refers to the full FOV of $5.5^\circ\times 5.5^\circ$ but
rather thin redshift slice to avoid large variation in the volume
of the resolution elements.
The smoothing effect induced by the volume averaging is very
evident in the poor sampling of the high-density and high-temperature
regions, when compared to Fig.~6 of \cite{2009ApJ...697..328B}.
The effect can be further appreciated by 
looking at the black curve in
the bottom panel of Fig.~\ref{fig:rhoTdetected},
which shows the probability distribution function of the
gas density mass-weighted within each resolution element.
The gas density on the $X$--axis is expressed in units of mean
density.
The distribution function, which was obtained by marking
the phase-space plot over the gas temperature, looks
remarkably symmetric. This symmetry is a result of the averaging process 
that deletes the high density tail that characterize the density 
distribution function of the gas particles \citep{2001ApJ...552..473D}.

The top right panel of Fig.~\ref{fig:rhoTdetected} is analogous to the top left 
one but considers only gas elements
for which the combined  \ion{O}{7} and \ion{O}{8} 
line surface brightness is above the 5 $\sigma$ detection threshold of
0.07~\phot,
line systems that could be detected with 
 1 Ms observation with CRIS. This plot clearly indicates that 
 next-generation instruments will preferentially probe the 
 hottest part of the WHIM with $T>10^{6}$~K with typical
 density $\deltag \sim 100$, hence confirming the results of 
 \cite{2010MNRAS.407..544B}.
 
This result is better illustrated by the red dotted curve in 
the bottom panel of Fig.~\ref{fig:rhoTdetected} which is analogous to the
black curve but only refer to the gas that can be potentially
detected in a 1 Ms observation.
It clearly shows that detectable emission lines
preferentially probe  high density  regions.
The integral of the 
 the red dotted curve shows that 30\% of the
total gas mass (the integral of the black curve)
could be probed via emission line spectroscopy. The remaining
70 \% of the gas that will go undetected
typically resides in regions with density
below 40 times the cosmic mean and constitutes the bulk of the WHIM.
Reducing the exposure time to 100~ks further decreases the fraction of 
detectable gas to $\sim 20$\% and shifts its mean density to larger values
(blue dashed).

To further investigate the issue of the WHIM detectability, we plot
in Fig.~\ref{fig:SBrho}  the contours of gas mass fraction as a function
of gas density ($X$--axis) and line surface brightness ($Y$--axis).
The
two plots refer to the  \ion{O}{7} triplet (left panel) and \ion{O}{8}
line (right panel), respectively,
and consider all gas elements with $T>10^5$~K as in the top-right 
panel of Fig.~\ref{fig:rhoTdetected}. The first thing to notice is the
very large range (about 10 order of magnitudes) of line surface brightness. 
Deep observations with a \XENIA like mission will be capable of probing
the gas above a detection limit of 
$\sim 0.1$~\phot.
The expected line surface brightness spans $\sim$2 order of magnitude
when $\deltag \gtrsim 100$.  Below this value, and for a line
surface brightness below the detection threshold, the iso-probability
contours run almost parallel to the $Y$--axis, indicating the difficulty of
detecting
gas in lower density environment: huge improvements in
the instrumental sensitivity would be required to probe the bulk of the WHIM in emission.

\subsection{Estimating the Gas Temperature}
\label{sec:tgas}

\begin{figure}
\epsscale{.80}
\centering
\includegraphics[height=0.98\columnwidth,angle=-90]{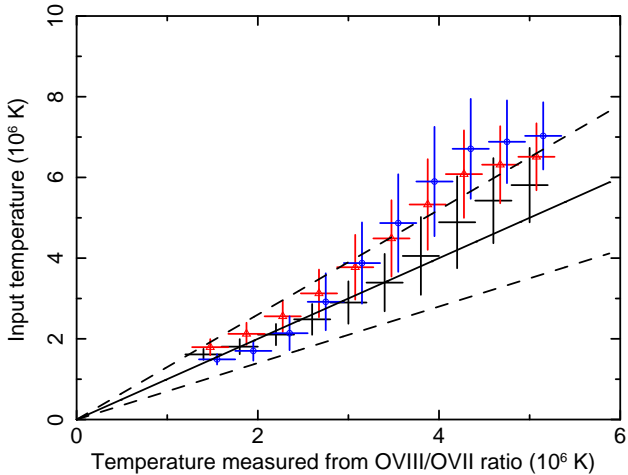}
\caption{
Relation between the gas temperature measured from the
 \ion{O}{8}-\ion{O}{7} line intensity ratio
in the mock WHIM spectra ($X$--axis) and different estimators for the underlying temperature of the 
gas ($Y$--axis). 
The $Y$--axis of the black, red, and blue crosses refer to 
\ion{O}{7}-emission weighted temperature,
\ion{O}{8}-emission weighted temperature, and
density weighted temperature, respectively.
The data are taken from
the FOV of $5.5^\circ\times5.5^\circ$ and the redshift range of
$0.202<z<0.274$.
The angular size of each volume element
$1.3'\times 1.3'$.
Horizontal error bars correspond to the temperature bin size, while
the vertical bars represent the scatter among resolution elements
in the temperature bin.
The perfect correlation (i.e., $Y=X$) is represented 
by the central solid line. 
The two external dashed lines represents $Y=1.3 X$ and $Y=X/1.3$,
respectively.
\label{fig:temp}}
\end{figure}

The quantitative analysis of the mock spectra allows us to investigate
the thermal properties of the emitting gas, namely to estimate the gas
temperature.  For those spectra in which both the \ion{O}{7} and \ion{O}{8} lines
have been detected one can use the line ratio to estimate the gas
temperature. Alternatively one could estimate the gas temperature 
by comparing the relative
intensity of the resonance, forbidden, and intercombination lines of the
\ion{O}{7} triplet. These, however,  can be detected only
 in the $\sim 20$ \% of the cases.
Therefore in this
section we assess  how well one can measure the gas
temperature from the \ion{O}{7}--\ion{O}{8} line intensity
ratio when the CIE hypothesis holds true.
For this purpose  we have
compared the temperature estimated from the mock spectra to the ``true''
temperatures of the gas responsible for line emission.
The latter quantity, however, is not uniquely defined for two reasons. 
First, there is no one-to-one correlation between the measured quantity, which 
is derived from the line intensity, and the temperature of the gas particles.
Second, the line emitting regions are sampled in finite volume elements 
because of the angular/energy resolution of the spectrometer.
To tackle this problem we will use different estimators of the ``true'' temperature
and compare the results with the temperature estimated from the
\ion{O}{7}--\ion{O}{8} line intensity ratio.

In Fig.~\ref{fig:temp} we have plotted the correlation between the
temperature measured from the mock spectra ($X$--axis) and 
the one obtained from three different gas temperature estimators
($Y$--axis). Here we consider the case of model B2 only.
In the plot the different colors indicate the three  different estimators:
\ion{O}{7}-line-emission weighted temperature (black crosses)
\ion{O}{8}-line-emission weighted temperature (red crosses)
mass-density weighted temperature (blue crosses).
The ``true''gas temperature is estimated in those 
resolution elements  where the surface
brightness of both \ion{O}{7} and \ion{O}{8} is above
0.07~\phot.
The vertical size of each cross represents 
the scatter about the mean. The horizontal bar represents 
the size of the corresponding temperature bin.

The plot shows that a tight correlation between the measured and the 
``true'' temperature is found
with all estimators used.  The temperature estimated from the
\ion{O}{8}--\ion{O}{7} line intensity ratio 
seems to best
trace the   \ion{O}{7}-emission weighted temperature of the gas, which is expected since 
temperature estimated from spectral lines preferentially  weight those regions in which
line emission is stronger.  
When the \ion{O}{8}-emission weighted temperature is used the correlation is still tight but the
slope is larger than unity. This is due to the fact that this estimator preferentially weights
regions in which the \ion{O}{8} is strong, i.e. regions in which the temperature is typically higher than
those from which the \ion{O}{7} photons come from.
With the mass-density weighted temperature the relation is unbiased for
temperatures
below  $T\sim 3\times10^{6}$~K and then the correlation becomes steeper.
This behavior can be understood as follows. The
 temperature measured from the spectra
is  sensitive to $T=$(2--3)$\times10^{6}$~K, where \ion{O}{7} and 
\ion{O}{8} emissivity has a peak and it is reliably traced by the 
mass weighted temperature. 
At higher temperatures,
a large fraction of the gas is too hot to contribute to the \ion{O}{7}
emission and therefore it is not traced by the \ion{O}{7}
line. As a result,
the measured temperature underestimates the mass-weighted
temperature of the gas.

Our assumption of CIE is well justified for gas in high 
density and temperature environments ($\delta_g\gtrsim 100$ and $T\gtrsim 2\times10^6$~K).
Photoionization becomes non-negligible in regions characterized 
by lower density and temperature (i.e. when 
$\delta_g\lesssim 100$ and $T\lesssim 2\times10^6$~K).
The presence of the photoionizing background would change the
ionization state of the gas.
In particular, this background would ionize \ion{O}{7} into \ion{O}{8}
\citep{2003ApJ...594...42C}, and 
thus enhancing  the intensity of the \ion{O}{8} emission line  
relative to that of \ion{O}{7}.
As a consequence, assuming the \ion{O}{7}/\ion{O}{8} ratio
of CIE overestimates the temperature of the gas. 
However, these effects will affect the gas properties in regions of
moderate overdensity and temperature which could hardly
be observed through emission line spectroscopy with current
and next-generation detectors.

These results show that one can use the measured \ion{O}{7} and
\ion{O}{8} 
lines to 
estimate the temperature of the gas with  a reasonable precision
(10--20\% uncertainty,
depending on the gas temperature). The measured quantity probes the temperature of
the gas responsible for \ion{O}{7} emission and underestimates that of the \ion{O}{8} emitting regions.
The measured quantity also traces the temperature of the bulk of the gas for 
$T=$(1--3)$\times10^{6}$~K and underestimates it in hotter regions.

Finally, we note that we have assumed a detector with
angular resolution $1.3'\times 1.3'$
and that we do not consider the 
line broadening due to the detector response.
The rationale behind this choice of the angular resolution
is to reduce possible systematics in the estimate of the ``true''
temperature.  In fact, using an angular resolution of 
$2.6'\times 2.6'$
as in the rest of the paper would not change appreciably the
value of the measured temperature but would artificially reduce the
estimated ``true''
temperature of the gas because of the increase in the
filtering volume associated to the resolution element.
A good angular resolution is thus important for 
an accurate determination of the physical state of the WHIM. 
Including the effect of the line broadening induced by the energy resolution of
the detector would reduce the number of detectable line systems and
would increase the statistical errors. However, we do not expect that
these errors would seriously affect our temperature estimates since
temperature errors are driven by the presence of multi-phase gas within
the resolution elements and not by the limited line statistics.

\section{Probing the spatial distribution of the WHIM}
\label{sec:3d}

Spatially resolved spectroscopy provides the unique possibility of tracing the 
spatial distribution of the WHIM. In this section we use our simulated 2D spectra to 
investigate the 3D distribution of the WHIM responsible for detectable line emission.
Here we focus our attention on the WHIM model B2 and assume a 1~Ms observation with 
a CRIS-like spectrometer with angular resolution of
$2.6'\times 2.6'$, energy 
resolution of 1~eV and effective collecting area of 1000~cm$^2$.

We only consider systems in which both \ion{O}{7} and
\ion{O}{8} lines can be detected at $\ge 5\sigma$ significance 
and ignore the possibility 
of using additional metal lines associated to the source.
The need of using both lines is dictated by the need of measuring the redshift of the 
associated line emitting region. This is guaranteed by selecting line pairs 
with energy ratio consistent with that of the \ion{O}{7} and
\ion{O}{8} lines.
Using additional metal lines can, in principle, improve the identification of line emission systems
and the determination of their redshift. However, the benefit in using these extra lines is in practice
limited by the associated increase  in chance contaminations.
For example, a number of  emission lines from Fe L complex
can be seen in the mock spectra. However, these lines typically probe the hotter part of the 
WHIM, typically associated with virialized structure also responsible for strong continuum 
emission which reduces the significance of line detections. 
The \ion{Ne}{9} is a better signpost for the WHIM since it probes 
gas with $T\sim 10^{6.6}$~K. However its emission line is too close to the Fe L lines to
eliminate the risk of spurious detection.
Finally, the energies of the redshifted N and C lines are typically found at the edges 
of the instrumental range, where 
a forest of emission lines due to L and M transition of heavier
metals exists.
For all these reasons we search the spectra for emission systems characterized 
by the \ion{O}{7} K$\alpha$ resonance and \ion{O}{8} Ly$\alpha$ pair and do not
look for additional ion lines.

Fig.~\ref{fig:mockmap1} and ~\ref{fig:mockmap2} compare the 3D 
distribution of the selected \ion{O}{7} + \ion{O}{8} line systems (lower 
panels) with the  spatial distribution of the gas in the simulation
(upper panels). Both figures show the same FOV of $5.5^\circ\times5.5^\circ$,
corresponding to a $6\times6$ mosaic with CRIS,  but refer to two different redshift slice.
Fig.~\ref{fig:mockmap1} refers to the redshift interval $z=[0.2117,0.2317]$ 
corresponding to a 54$h^{-1}$Mpc slice at a comoving distance of 630 $h^{-1}$Mpc
in the cosmological model of the hydrodynamical simulation.
Fig.~\ref{fig:mockmap2} shows a much closer slice of 57$h^{-1}$Mpc 
at a comoving distance of 265 $h^{-1}$Mpc, corresponding to
$z=[0.0805,0.1004]$.

\begin{figure}[!t]
\centering
\includegraphics[width=0.98\columnwidth]{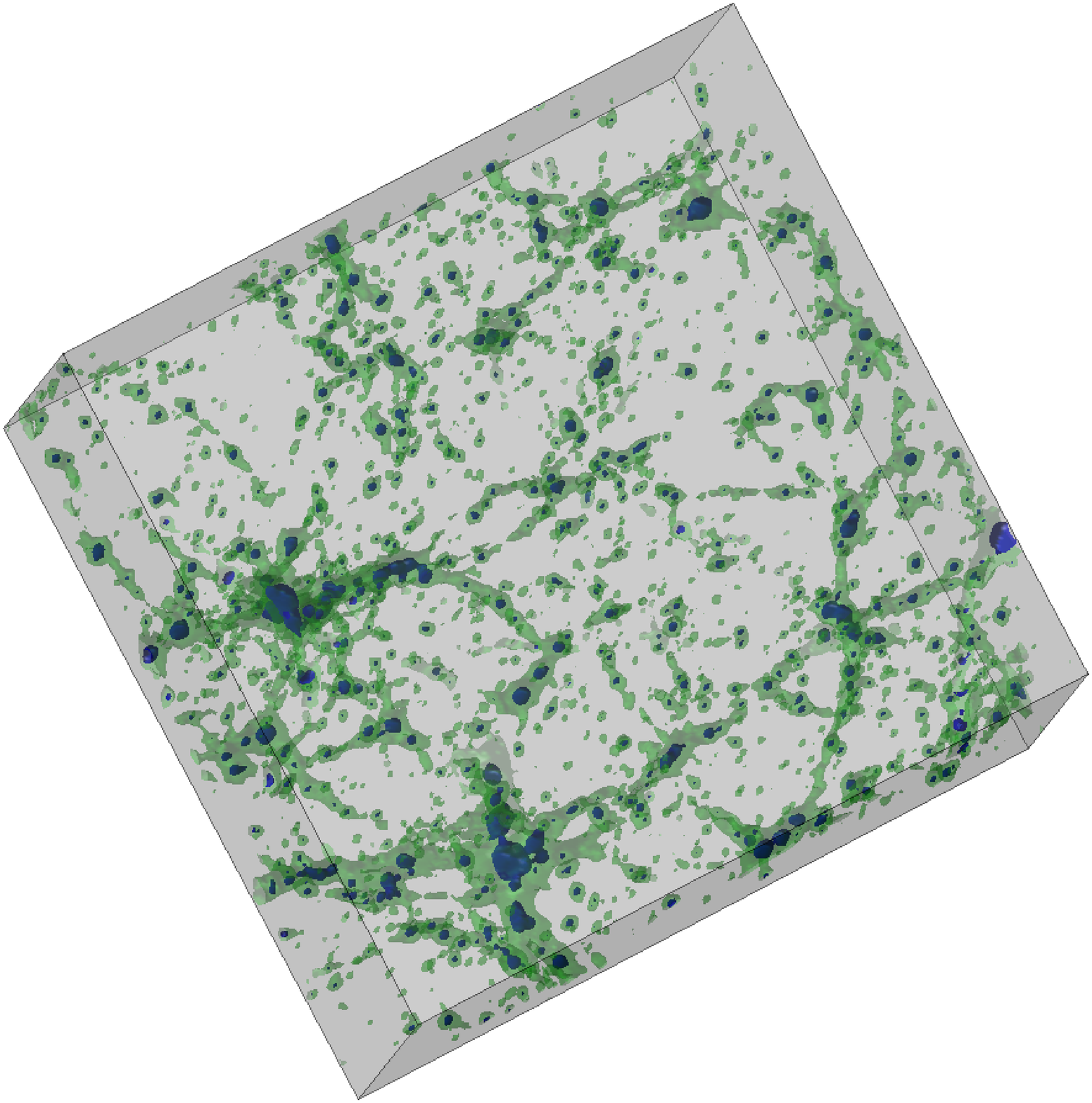}
\includegraphics[width=0.98\columnwidth]{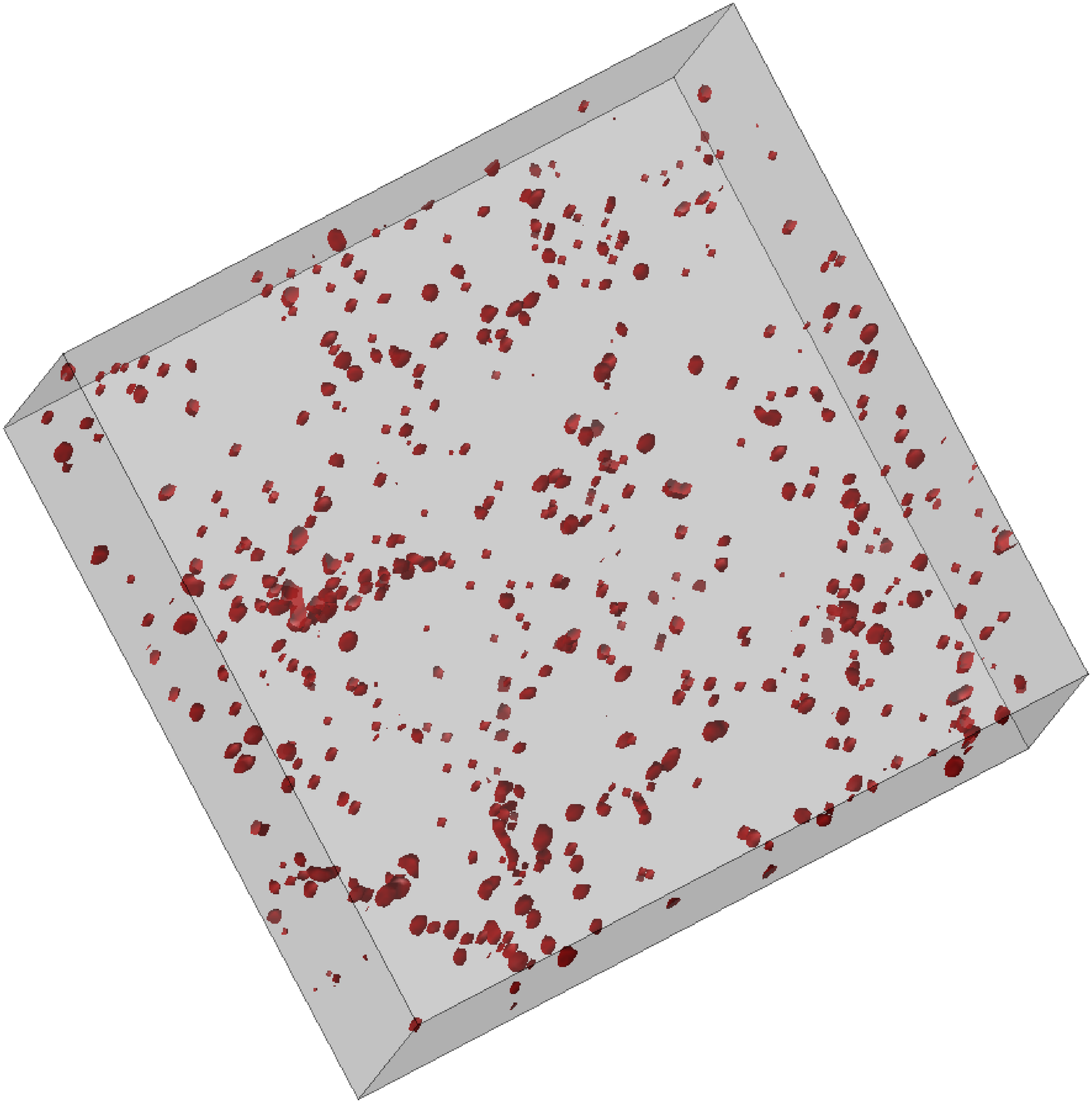}
\caption{ 
Slice of gas density contour (top) and O emission map (bottom)
at $z=0.2117-0.2317$, corresponding
to comoving distance of 603--657$h^{-1}$ Mpc.
The field of view is $5.5^\circ\times5.5^\circ$.
The blue region of the top panel corresponds to
$\rho_\mathrm{gas}/\left<\rho_\mathrm{gas}\right>=75$, while
the green region corresponds to 
$\rho_\mathrm{gas}/\left<\rho_\mathrm{gas}\right>=10$.
The red regions (blobs) show where both
\ion{O}{7} and \ion{O}{8} lines are identified with $>5\sigma$
significance, with 1~Ms exposure by {\it Xenia} CRIS-like
instrument.
\label{fig:mockmap1}
}
\end{figure}

\begin{figure}[!t]
\centering
\includegraphics[width=0.98\columnwidth]{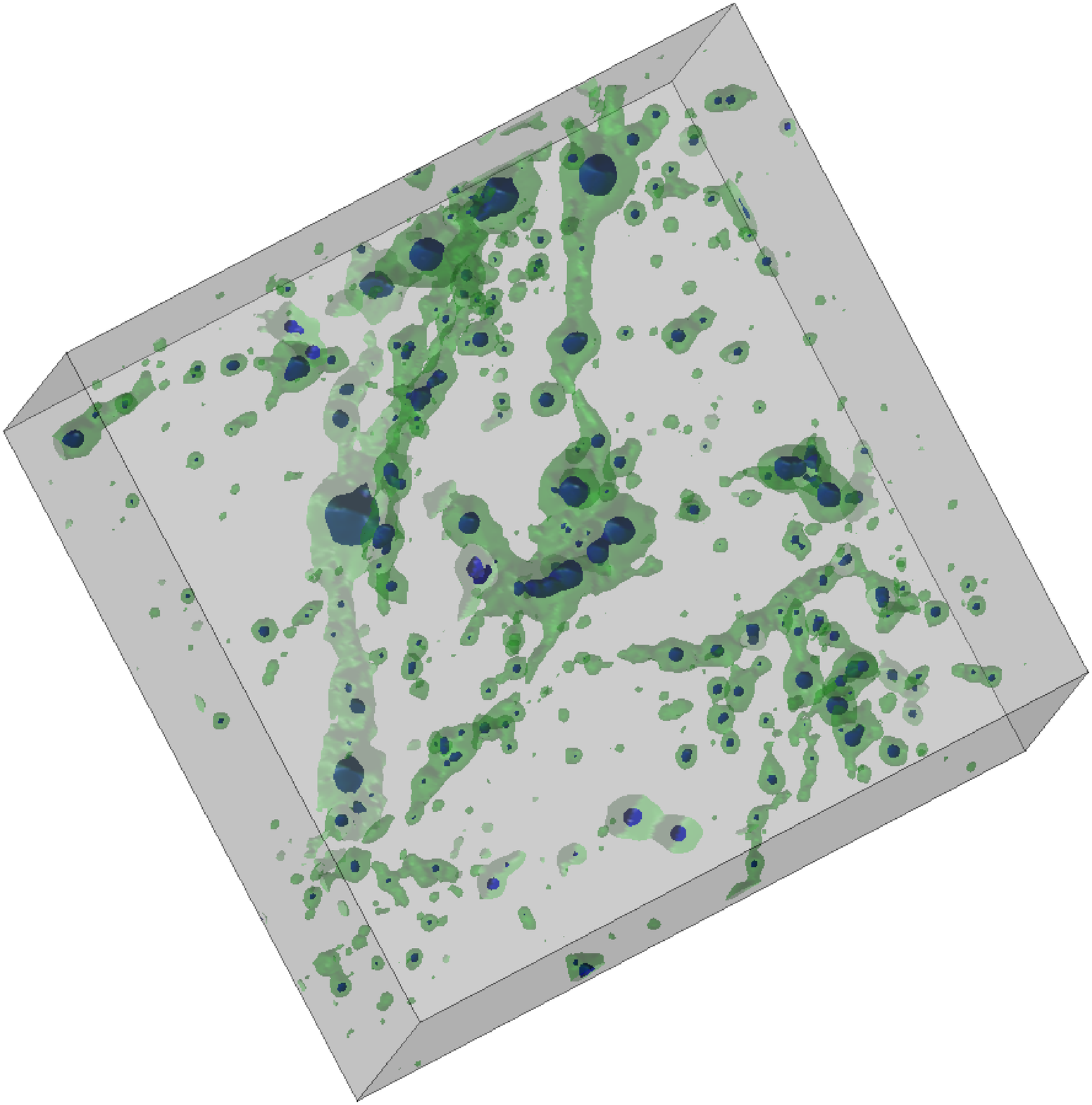}
\includegraphics[width=0.98\columnwidth]{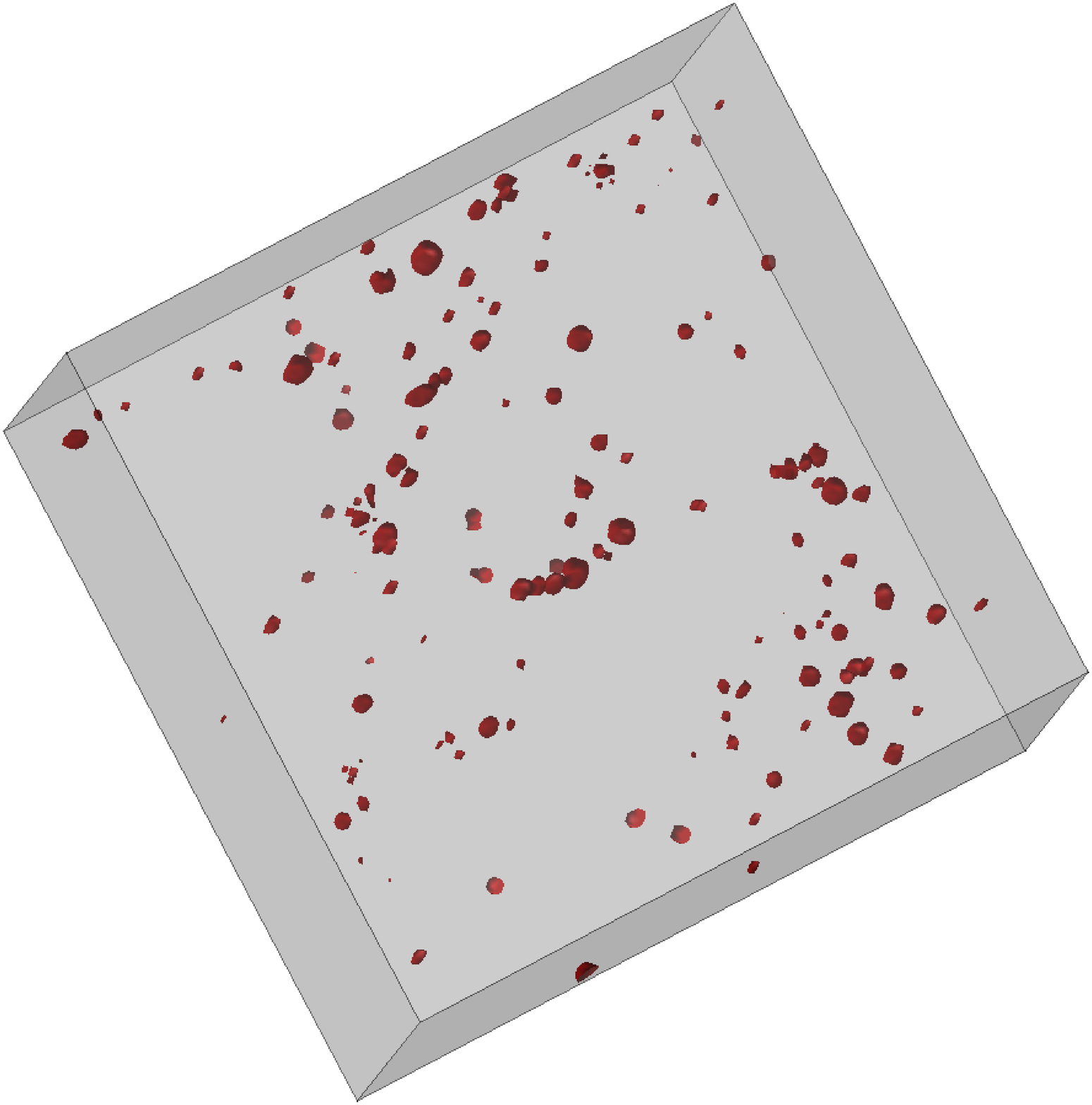}
\caption{ 
Same as Figure~\ref{fig:mockmap1}, but 
slice at $z=0.0805-0.1004$, corresponding
to comoving distance of 237--294$h^{-1}$ Mpc.
The field of view is $5.5^\circ\times5.5^\circ$.
\label{fig:mockmap2}
}
\end{figure}

In the gas map the blue regions correspond to iso-density surfaces
drawn at $\rho_\mathrm{gas}/\left<\rho_\mathrm{gas}\right>=75$. 
The green surfaces corresponds to regions characterized by 
smaller overdensity $\rho_\mathrm{gas}/\left<\rho_\mathrm{gas}\right>=10$.
At these overdensity levels the spatial distribution of the gas in the simulation is 
characterized by the well-known network of filaments punctuated by virialized 
structures that, in this iso-contour maps, appear as quasi-spherical 
``blobs''.
The filamentary structure is more clearly seen in the more distant slice thanks to 
the large volume sampled. The gas distribution in the nearby slice is clearly more 
dominated by a few, prominent structures, as expected.

The red regions in the bottom panels identify the positions of the 
selected \ion{O}{7} + \ion{O}{8} line systems. Clearly these line systems 
qualitatively trace, in a sparse fashion,  the large scale structures
 of the gas 
with overdensity  $\rho_\mathrm{gas}/\left<\rho_\mathrm{gas}\right>=75$.
However the tracing is not unbiased since   the line systems
tend to undersample the more massive virial structures associated to the 
largest blob. This bias reflects the unfavorable ion balance for
\ion{O}{7}  and \ion{O}{8} within hot virialized regions.
The fact that  \ion{O}{7}$+$\ion{O}{8} line systems are poor tracers of 
gas at lower overdensity ($\rho_\mathrm{gas}/\left<\rho_\mathrm{gas}\right>=10$)
confirms the fact that emission lines preferentially samples the 
fraction of the WHIM that resides in high density regions
(Fig.~\ref{fig:rhoTdetected}).

This qualitative analysis confirms that next-generation X-ray spectrometers 
could be successfully used to trace the spatial distribution of the 
WHIM through its line emission. 
Thanks to the large number of the detected emission line systems,
more quantitative analysis can be performed
through statistical estimators.
In \cite{ACFpaper}, which uses the same mock spectra,
it is shown that the two-point spatial correlation function of
the line-emitters could be measured with a good accuracy.
Its slope and correlation length, which should be similar to
those of the galaxies, could be estimated at a few \% accuracy level.

The tracing of underlying gas by the detected line systems, which in the
plots looks quite patchy, is possibly improved by adopting some adaptive
smoothing technique like those based on the Wavelet transform
analysis. The improvement would be particularly noticeable for nearby
structures since, in this cases, the angular size of the associated gas
cloud is larger than the angular resolution of the instrument.

\section{Robustness tests against non-ideal effects}
\label{SEC:cons-non-ideal}

In this section we assess the impact of foreground emission, energy
resolution and contamination from spurious metal lines on the
detectability of the extragalactic \ion{O}{7} and 
\ion{O}{8} lines and on our
capability of tracing the spatial distribution of the WHIM.

\subsection{Impact of contamination from spurious lines and
high continuum}
\label{SEC:detect-emiss-source}

We address the problem of identifying \ion{O}{7} and \ion{O}{8} lines
with no {\it a priori} knowledge of the nature of the detected lines and of
their redshift.  For this purpose the line-detection
procedure have been modified from that depicted in \S~\ref{sec:linedet}.
The rationale behind this choice is to avoid contamination from 
lines associated to different line emitting regions along the line of sight and to account for the 
fact that the line detection efficiency is not constant across the FOV  since 
the level of the continuum emission depends on the direction.
We have decided to ignore additional metal lines associated to the WHIM
since their search would increase the chance of false identifications.
To identify the putative WHIM emission systems we do not
perform spectral fitting assuming some particular model 
but we search for the \ion{O}{7} K$\alpha$  
and \ion{O}{8} Ly$\alpha$ line
pairs with a strategy
more conservative than that adopted in  \S~\ref{sec:detection}
and designed to account for the variable spectral continuum and for the presence 
of spurious spectral features. The procedure is as follows.
{\it (1)} We deal with the fact that the continuum level may vary along the spectrum and
also from spectra to spectra,  across the FOV.
To account for this effect we modify the line detection criterion: instead of  setting a fixed 
line detection threshold, corresponding to a minimum line surface brightness, we define 
the detection significance $S$ by comparing the number of photon counts
in 1~eV bin
({\it Counts})
 with the number of continuum counts ({\it Continuum})
obtained by averaging over the neighboring 50~eV that do not contain line features above
2 $\sigma$ or known galactic lines brighter than 
$>0.05$~\photeV.
The significance is then defined as:
\begin{equation}
S/N = \frac{\mathit{Counts}-\mathit{Continuum}}
 {1.0+\sqrt{\mathit{Continuum}+0.75}}
\label{eq:sigmaObs}
\end{equation}
where the denominator is an approximate
 1$\sigma$ Poisson error 
\citep{1986ApJ...303..336G}.
Note that Eq.~(\ref{eq:sigmaObs}) is a similar quantity
as Eq.~(\ref{eq:sn}) but is a quantity calculated with observed
quantities.
{\it (2)} We consider only lines in the energy range [375,653]~eV
detected at 5$\sigma$ significance (i.e., $S/N>5$). 
With this criterion we identify
$7.8\times10^3~ \mathrm{lines~ deg^{-2}}$
corresponding to transitions of highly ionized ions of N, O, Ne,
and Fe.
{\it (3)}
We minimize the chance of contamination by spurious lines by enforcing a
number of constraints.
 {\it (i)} we reject all lines with flux of 
$>0.05$~\photeV\ 
 that could be attributed to Galaxy emission, i.e., all known Galactic lines;
 {\it (ii)} to avoid oversampling the same structures we discard all lines with energy
within  $15$~eV from that of a stronger emission line;
 {\it (iii)} we search for line pairs from the same angular
element, and pick up the pairs only if the 
energy interval of the two lines is the same as that of
redshifted \ion{O}{7} K$\alpha$ resonance and \ion{O}{8} Ly$\alpha$
lines.  The emission lines without an associated \ion{O}{7} or \ion{O}{8}
line are discarded.

Using the procedure, we only consider those systems that both the \ion{O}{7} K$\alpha$
and the \ion{O}{8} Ly$\alpha$ lines are identified with significance larger than
5~$\sigma$.
As a result we obtain 446~deg$^{-2}$  line emission systems identified.
This number is $\sim30\%$ less than those identified in \S~\ref{sec:detection}.
The discrepancy comes from the combined effect of line contamination and 
high continuum level.
We confirm that, with this procedure, the contamination of 
chance coincidence of
either \ion{O}{7} or \ion{O}{8} lines 
with heavier metal lines is less than 10\%.
Our spectra do not include photons emitted by gas at $z>0.5$.
\ion{O}{7} lines
emitted beyond this redshift would be 
redshifted out of the energy range probed by the detector, so the only 
impact of high redshift gas would be to increase the probability of 
contamination by  higher energy lines such as \ion{O}{8},
\ion{Ne}{9}, \ion{O}{10} and Fe L complex. However,
we do not expect a severe additional contamination for two reasons.
First, the dimming of the emission line intensity, which scales as 
$(1+z)^{-3}$, is not exactly counterbalanced by the 
increase in the probed  volume since,the latter decreases 
more slowly than $\propto (1+z)^{3}$ at high redshifts 
Second, at higher redshifts the mass fraction of the 
gas in the Ly-$\alpha$ forest increases while  the 
abundance of the gas contributing to  \ion{O}{7}
and  \ion{O}{8} lines emission decreases.
To test the validity of these hypotheses and quantify 
line contamination by high-redshift gas 
we have created and analyzed mock spectra 
that accounts for emission by the gas out to
$z<1.2$ and within a FOV of $2.6^\circ\times 2.6^\circ$
and have compared the results with the reference case
of $z<0.5$. Including the emission from gas beyond
$z=0.5$ decreases the number of detectable 
\ion{O}{7}--\ion{O}{8} pairs  by only 10--15\%..

\subsection{Impact of energy resolution}
\label{SEC:infl-finite-energy}

A degradation in the energy resolution of the instruments affects the
possibility of detecting the WHIM lines in two ways. First, it
decreases the signal-to-noise ratio in Eq.~(\ref{eq:sn}).
As a consequence, one needs
to increase the surface brightness threshold for $5\sigma$ detections,
hence reducing the total number of detected lines. Second, with a large
$\Delta E$ the chance of contamination from neighboring lines
increases. Here we let the energy resolution vary in the range [1,7]~eV
which encompasses the values already achieved with current
instrument \citep[$\Delta E=7$~eV with
{\it Suzaku};][]{2007PASJ...59S..77K}
and the energy
resolutions expected by currently planned X-ray missions equipped with
a microcalorimeter spectrograph.

A visual impression of the cumulative impact of these two effects is
given by Fig.~\ref{fig:DEinfluence},
which shows a portion of the mock emission spectrum
displayed in Fig.~\ref{fig:emspec}
but observed with different energy resolutions and in log scale.
With 1~eV resolution (FWHM), \ion{O}{7} triplets and \ion{O}{8}
resonant line at $z=0.033$ (0.55 keV and 0.63 keV, respectively) are
clearly detected, with a marginal detection of a emission line at 0.47
keV.  
With $\Delta E=3$~eV, the 0.47~keV line is below
detection threshold, and with $\Delta E=5$~eV
the \ion{O}{7} triplet becomes hard to
observe because of the local ($z=0$) 
\ion{O}{7} emission. When $\Delta E$ is 7~eV,
the \ion{O}{7} triplet drops below detection threshold.

\begin{figure}[!ht]
\centering
\includegraphics[height=0.98\columnwidth,angle=-90]{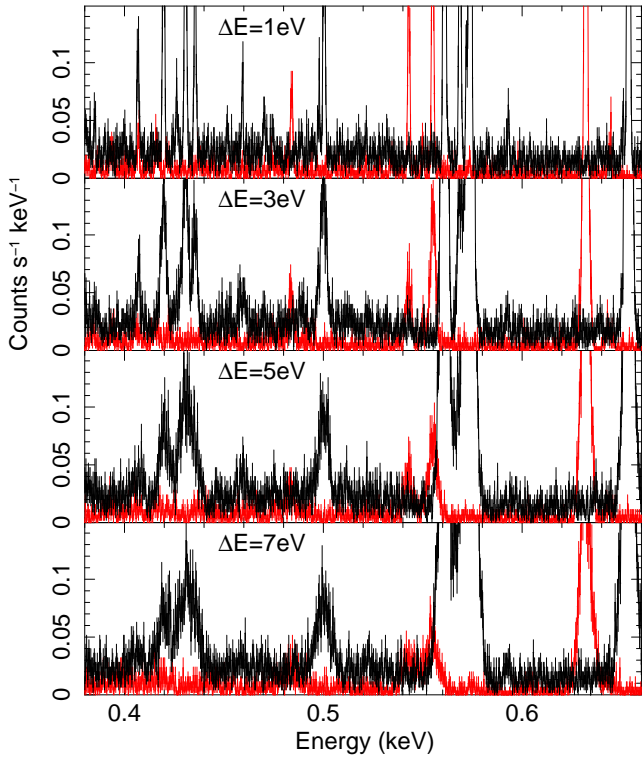}
\caption{ 
Mock spectrum in a $2.6'\times 2.6'$  area from a 1~Ms
exposure with {\it Xenia} CRIS.
The black spectra are the sum of Galactic foreground and 
unresolved extragalactic  background emission, while the red is
extragalactic diffuse gas.
The four panels of the each plot show the spectra convolved
with the energy resolution (FWHM)
of 1~eV, 3~eV, 5~eV, and 7~eV, respectively.
\label{fig:DEinfluence}
}
\end{figure}

\begin{deluxetable*}{ccccccc}
\tabletypesize{\small}
\tablecaption{Number of emission line detections with
different detector energy resolution
\label{tab:infludenceDE}
}  
\tablewidth{0pt}
\tablehead{
\colhead{$\Delta E$}  &
\colhead{$f_{\rm line}$} &
\colhead{$dN_{\rm OVII+OVIII}/dz$} &
\colhead{Fraction of} &
\colhead{$N_{\rm OVII+OVIII}$  per deg$^2$}\\
(eV) & (\phot)
& \nodata & high $f_{\rm FG}$ (\%) & in the simulation
}
\startdata
---\tablenotemark{a} & 0.07 & 2.4 & 14 & 446\\
1 & 0.07 & 2.4 & 20 & 396\\
3 & 0.11 & 2.1 & 30 & 351\\
5 & 0.13 & 2.0 & 36 & 274\\
7 & 0.15 & 1.9 & 41 & 163\\
\enddata
\tablecomments{
Column 1: Detector energy resolution.
Column 2: Minimum line surface brightness required for a 5$\sigma$ detection
(\phot).
Column 3: Expected number of simultaneous  \ion{O}{8} and \ion{O}{8} detections 
per resolution element and unit redshift.  The foreground emission
is assumed to have a constant surface brightness of
20~\photkeV.
Column 4: Fraction of the energy range where foreground emission
surface brightness exceeds 
20~\photkeV.
Column 5: Number of simultaneous \ion{O}{8} and \ion{O}{8} detections 
per square degree in the mock spectra.
All estimates assume an angular resolution of 
$2.6'\times 2.6'$ 
and model B2.
}
\tablenotetext{a}{Not convolved with detector energy resolution.
The bin size is 1~eV.}
\end{deluxetable*}

A quantitative assessment of the contamination by spurious lines is
provided by Fig.~\ref{fig:GalacticLineSB}.
The fraction of energy bins with surface
brightness above that of the Galactic foreground emission
 ($f_\mathrm{FG}=20$~\photkeV)
steadily increases with 
$\DE$: from 20\% for $\DE=1$~eV to
41\% for $\DE=7$~eV.  
In Table~\ref{tab:infludenceDE}
 we quantify how energy resolution affects the
number of expected detections, splitting its effect in two: the increase
of the surface brightness detection threshold (column 3) and the
increase in the number of bins contaminated by the Galactic Foreground
(column 4).  All estimates assume the same setup as in 
Table~\ref{tab:tab2}
apart from energy resolution: WHIM model B2, 1~Ms observation, and 
$2.6'\times2.6'$
angular resolution. The first effect is rather minor and reduced
the expected number of line detection only by $<20\%$. On the other hand,
contamination by galactic foreground has a more serious impact since the
number of expected detection is directly proportional to that of the
uncontaminated energy bins. Our results show that decreasing $\DE$ from 1 to 
7~eV reduces the number of expected detection by a factor of 2.  It should
also be noticed that this effect does not randomly affect all emission
line systems.  The systems that are more seriously affected are those
with redshifted emission lines that coincide with those of the Galactic
foreground. This induces a non-trivial selection effect in the
redshift distribution of the WHIM emitters that need to be corrected for
to trace their 3D distribution.

To quantify the impact of the energy resolution on the expected number
of WHIM line detections we have repeated the analysis performed in
\S~\ref{SEC:detect-emiss-source}
after convolving the mock spectra with the appropriate
response matrix.  
In the analysis of the response-convolved spectra,
we again do not fit the 
spectra with a particular model, but searched for the \ion{O}{7}
and \ion{O}{8} line pairs.
The only difference is that in the new analysis
photon counts in neighboring energy bins are summed up before searching
for emission lines. 
This is done to collect photons that are spread into several 
energy bins (the bin width is 1~eV)
due to the energy resolution of the instrument.  
The results of this procedure obviously
depend on the number of neighboring bins to be summed upon. We set this
free parameter by maximizing the number of detectable emission lines.
It turns out that the best results are obtained when summing over
(($\DE/1~\mathrm{eV})+1$) bins for $\Delta E \le 5$ eV and over 
(($\Delta E/1~\mathrm{eV})-3$) for $\Delta E = 7$ eV. 
The change of the trend for $\Delta E =7$~eV 
reflects the fact that at low energy resolution foreground
contamination becomes the limiting factor.  
The number of expected
detections is listed in column 5 of Table~\ref{tab:infludenceDE}. 
It is clear that an energy resolution
$\Delta E \le 3$ eV is desirable, although even with $\Delta E = 7$ eV
one still expects 160 \ion{O}{7}$+$\ion{O}{8} line 
detections per square degree.  The
reason why the number of expected detection of the `no convolution' case
is different from the analogous one listed in Table~\ref{tab:tab2}
is because in the
line detection strategy used in this case, and described in
\S~\ref{SEC:detect-emiss-source},
we have ignored the energy range in which the surface brightness of the
foreground emission is above $f_\mathrm{FG}$. These ranges could be also be
searched for detectable lines by locally increasing the detection
threshold. However the expected increase in the number of detections is
small since the intensity of the foreground emission lines is typically
an order of magnitude larger than that of the WHIM lines.

\section{Discriminating among WHIM models}
\label{SEC:capab-dist-diff}

\begin{figure}[!htb]
\centering
\includegraphics[height=0.98\columnwidth, angle=-90]{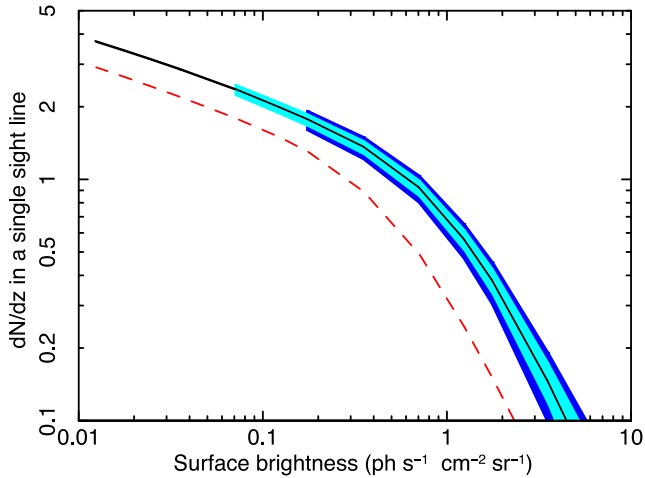}
\caption{
The expected $dN/dz$ curve with a Poisson error with 1~Ms exposure,
$2.6'\times2.6'$ angular size, and 
1 deg$^2$ FOV with CRIS.  The black curve is the model
curve (model B2). Color-filled areas are 1$\sigma$ Poisson
error regions: cyan and blue are without considering detector energy
resolution and with convolution with 7~eV energy resolution,
respectively.
Red dashed line is from model B1 (Poisson error is not shown).
\label{fig:dNdzPoisson}
}
\end{figure}

As discussed in \S~\ref{sec:whimodel}, current limitations in our
understanding of the stellar feedback, chemical evolution, and metal
diffusion process are reflected in the large uncertainties of the WHIM
models.  The most recent hydro-dynamical simulations of the
intergalactic gas at low redshifts have been used to construct different
WHIM models by assuming different, physically plausible, recipes for the
above processes. Comparisons between model predictions and observations
can therefore constrain the 
cosmic history of 
chemical evolution and metal diffusion
along with the underlying stellar feedback processes by
discriminating among these WHIM models.  UV observations are
leading the way, since they currently provide the most stringent
constraints on the WHIM models through \ion{O}{6} line statistics.
X-ray observations can add additional constraints, through line
statistics, direct assessment on the physical state and spatial
distribution of the emitting gas.

In this section we assess to what extent future experiments will enable
us to discriminate among different WHIM models. Here we focus on the
$dN/dz$ statistics of \ion{O}{7} and \ion{O}{8} lines.
Fig.~\ref{fig:dNdzPoisson} shows the expected
number of \ion{O}{7}$+$\ion{O}{8} detection per unit redshift in an angular
resolution element of $2.6'\times2.6'$ and for 1~Ms observation with CRIS for
model B2 (black, continuous curve) and B1 (red, dashed). The
cyan-colored band around the black lines show the $1\sigma$ Poisson
uncertainties for model B2 
from the finite number of detections expected in a
$1~\mathrm{deg}^2$ FOV.  
The expected errors are much smaller than the difference
between model B1 and B2, suggesting that observations will 
efficiently discriminate among different WHIM models. The capability of
constraining the physical processes that regulate the WHIM is further
illustrated by pointing out that difference between model B1 and model
B2 is of the same order as that between the WHIM models proposed 
by \cite{2006ApJ...650..573C},
which adopt different prescription for the ionization
balance and for galactic superwinds.  The larger, blue uncertainty strips
that extend beyond the surface brightness threshold of 
0.15~\phot\
refers to the case of $\Delta E=7$ eV.
It shows that even
with a rather modest energy resolution the discriminatory power of the
$dN/dz$ test is still considerably large. 
The significance of discriminating the two models is
3--5$\sigma$ with a 1~deg$^2$ observation, 
depending on the energy resolution. 

It should be noted that the cosmic variance introduces an additional
error in determining the $dN/dz$ curve.  
We have estimated the contribution of cosmic variance 
by splitting the simulated $5.5^\circ\times5.5^\circ$ field 
into several smaller regions.  The variance 
in detectable emission line systems
is 20\%, 15\%, and 10\%,
for the FOV of 0.25~deg$^2$, 1.0~deg$^2$, and 4.0~deg$^2$,
respectively.  Hence, a large-field mapping with $>4.0$~deg$^2$ is desirable
to make the uncertainty due to cosmic variance smaller than
the statistical error ($\sim$10\% for 7~eV energy resolution).
This can be easily achieved with a dedicated mission such as
{\it EDGE} and {\it Xenia} thanks to  the large FOV.

Finally, we point out that emission-line studies provide better
statistics than absorption studies because of the comparatively larger
number of line detections.  In fact, the two approaches are very
complementary.  Emission studies preferentially probe regions of
enhanced density where the impact of stellar feedback is larger and
chemical enrichment is presumably stronger. On the contrary, absorption
studies preferentially probe regions of moderate overdensity where the
bulk of the WHIM resides and thus are best indicated for closing the
baryon budget at $z=0$. Space missions like {\it EDGE}, 
{\it Xenia}, and {\it ORIGIN} are
designed to perform both emission and absorption analyses 
thanks to their capability of fast
re-pointing that will enable us to use Gamma-ray bursts as distant
beacon and search for the characteristic absorption features of the
intervening WHIM \citep{2009ApJ...697..328B}.

Constraining abundance and physical properties of the 
WHIM from observations it is not straightforward. The first
reason is that
in order to trace the thermal history of the WHIM it requires an 
accurate and self-consistent treatment of 
metal line cooling which, instead, is ignored in the present 
work. The second reason is
the approximate treatment of 
a number of processes and quantities 
\cite[galactic winds,  AGN feedback, stellar 
IMF and sub-resolution turbulence; see][]{2010MNRAS.407..544B,2009MNRAS.395.1875O,2010arXiv1009.0261S,2010MNRAS.402.1536S}
 that may affect the thermal state of 
the WHIM in the current numerical experiments. 
Some of these processes have little impact on 
observable quantities such as \ion{O}{8}
surface brightness \cite[e.g.,][]{2010MNRAS.407..544B},
and hence it cannot be 
used to discriminate among competing WHIM models. On the other hand, 
emission line statistics and analysis will no doubt help to lift some of the 
model degeneracy, especially when coupled to observational constraints on
the spatial distribution of the line-emitting gas, like those derived from the 
estimate of the angular and spatial two-point correlation function of the O-line emission
\citep{ACFpaper}.

\section{Discussion and Conclusions}
\label{sec:disc}

In this work we have  analyzed the mock X-ray spectra extracted from the
the hydrodynamical simulation of \cite{2004MNRAS.348.1078B}
to investigate the possibility of detecting the WHIM in emission
with next-generation X-ray satellites,
assess its thermal state and trace its spatial distribution.
Our theoretical predictions are based on the WHIM model of 
 \citet{2009ApJ...697..328B}.
The theoretical predictions presented in this work are mainly based on their model ``B2''
 although we also consider the more conservative (and less realistic) WHIM model
 ``B1'' to bracket theoretical uncertainties. Finally, the experimental setup considered here 
 is the same one proposed for the {\it Xenia} mission.
 In particular we will assume that our mock spectra were ``observed''
 with the imaging spectrometer CRIS.

We showed that a CRIS-like spectrograph will enable us to detect a large
number of \ion{O}{7}$+$\ion{O}{8} line emitting systems, to estimate their
temperature from the line ratio and to trace their spatial distribution.
The line systems are expected to trace the hotter and denser part of the WHIM,
typically associated to the outskirts of large virialized structures.
Our results suggest that the number of expected detections is large enough to
discriminate among different WHIM models and to constrain the physical
processes that regulates its thermal state and metal content, even when
one accounts for contamination effects and allows for a moderate 
($\sim$7 eV)
energy resolution.

All the results presented in this work are based on a WHIM model that
fulfill all observational constraints but is nevertheless poorly
constrained by observations.  From the theoretical side, WHIM models
suffer from the limited knowledge of the feedback mechanisms and metal
diffusion processes that potentially affect the WHIM properties.  In
particular, the WHIM model adopted here, though effective in describing
observations, lacks self-consistency since the metallicity of the gas is
assigned in the {\it post-processing} phase rather than being computed
during the run. 
To fix these problems one should modify theoretical prescriptions for
feedback process or metal diffusion and gas cooling.  
Let us asses qualitatively the impact of having ignored line cooling.
Efficient metal cooling requires a cooling time $\tcool$  shorter than
the Hubble time $\tH$. With no metal cooling, $\tcool < \tH$ only when
$\delta_g > 1000$.
Cooling become more efficient in presence of metals and less efficient with the 
gas temperature, i.e. 
$\tcool$ decreases with metallicity  \cite[e.g., Figure~4 of][]{2009MNRAS.393...99W},
and increases with $T$.
On the other hand, 
we have shown that emission line spectroscopy 
can only detect gas with $T>10^6$~K (Figure~\ref{fig:rhoTdetected}).
Taking all this into account and assuming a reference metallicity 
of $\sim 0.1~Z_\odot$ one finds that $\tcool<\tH$ if 
$\delta_g>300$ for $T=10^6$~K and 
if 
$\delta_g>1000$ for $T=2\times10^6$~K. 
This simple arguments indicate that metal line cooling preferentially affects 
dense ($\delta_g>300$) gas typically associated to galaxy groups 
rather than the WHIM. Indeed, inefficient cooling in the simulation is 
probably the cause of the excess X-ray emission from groups in the simulation.

In fact, the impact of the metal cooling 
is likely to be more significant because it affects the
whole thermal history of the baryons
\citep{2009MNRAS.393...99W,2009MNRAS.395.1875O,2010arXiv1009.0261S,
2011MNRAS.tmp..165T}.
When metals are transferred from galaxies, either by galactic
wind or AGN, both the density and metallicity are high,
and hence metal cooling is very efficient.  After the temperature
is lowered by metal cooling, the gas is further
cooled down by adiabatic expansion.  As a result, the thermal state 
of baryons at $z\sim0$
is determined by their thermal history.  
The impact of these effects can only be assessed by 
running a numerical simulation in which metal cooling is 
accounted for self-consistently and,
more importantly, in which the relevant physical process 
are included correctly.
However, current numerical experiments which include a self consistent 
treatment of metal cooling  \citep{2010MNRAS.407..544B}
seem to corroborate our qualitative 
argument since they show that metal cooling mostly affect 
regions with  overdensity typical of galaxy groups, rather than the WHIM.
For this reason, we are quite confident that considering predictions 
from both  model B2 and the very conservative model B1 will
allow us to bracket theoretical uncertainties, including incorrect treatment of 
metal cooling in the numerical simulation.

\cite{2010MNRAS.407..544B} have shown that predictions on the
detectability of the \ion{O}{8} line are are quite robust to the
different model prescription with the exception of the cooling function.
They show that using a metal-dependent cooling instead of a
metal-independent scheme (as in our model) decreases by up to an order
of magnitude the SB of the \ion{O}{8} line in regions populated by {\it
non}-WHIM, but does not significantly alter the line emission
properties in regions of lower overdensity where the WHIM typically
resides.  
Reducing the artificially large brightness of large
density regions would decrease the noise level and increase the
probability of detecting the emission line systems associated to the
WHIM.  

From these considerations,
we conclude that our WHIM models are adequate to assess
the detectability through \ion{O}{7}/\ion{O}{8}
line pairs
of the WHIM in emission and are confident that
theoretical  uncertainties are effectively bracketed, perhaps too
generously, in the scatter among B1 and B2 model predictions.
On the other hand, the gas distribution in the phase-space
(Figure~\ref{fig:rhoTdetected}) and in the 3D space
(Figure~\ref{fig:mockmap1} and
\ref{fig:mockmap2}), especially in regions of high density and
moderate temperature,
 might not be accurate because of the
approximate treatment of gas cooling in the simulations.

The results presented in this paper assume collisional ionization
equilibrium.  This is a reasonable hypothesis since large part of
\ion{O}{7} and \ion{O}{8} lines that next-generation instruments are
expected to detect are likely to be produced by warm-hot gas in high
density environments where the CIE is a fairly good assumption
($\deltag\gtrsim 100$ and $T\gtrsim2\times10^{6}$~K).
The validity of the CIE hypothesis can be directly checked by analyzing 
a number of spectral features.
One is the relative comparison of the
intensity of the resonant, forbidden and intercombination lines of the
\ion{O}{7} triplet, which also provide an estimate of the gas temperature,
alternative to that obtained from the \ion{O}{8}/\ion{O}{7} ratio
 \citep[see, e.g.,][]{2001A&A...376.1113P}. 
The possibility of performing this test, however, is limited to the few cases in which the 
components of the triplets can be unambiguously detected. 
>From the analysis of the mock spectra, 
only in 30 \% of the cases can we detect both
the resonant and the forbidden lines.
The weaker intercombination lines is
only detected for  $\sim 10 $ \%.
If photo-ionization plays a role, \ion{O}{8} line is also emitted
from lower temperature gas, in addition to \ion{O}{7} triplets.
Therefore, the number of detected line systems would increase
because lower density ($\delta\lesssim 200$) regions becomes
also detectable.
The ionization state of the gas in this low density regions
would be affected and the temperature would be overestimated
under the assumption of CIE.

Our analysis of the spatial distribution of the line emission systems
presented in \S~\ref{sec:3d} is rather qualitative.  A more quantitative
analysis, 
based on the mock spectra described in this work,
is presented in \cite{ACFpaper} in which it is shown that the
number density of detectable emission line systems is large enough to
characterize the distribution by means of the spatial two-point
correlation function.  
It is also shown that the two-point correlation
function of the WHIM is significantly different from that of the {\it
non-}WHIM and similar to that of galaxies, to which the line-emitting
WHIM is associated.  Its correlation properties do not seem to evolve
significantly in the redshift range $z=[0,0.5]$ and the dynamical state,
induced from the velocity-driven anisotropies in the observed
clustering, is consistent with that of a gas coherently falling toward
larger structures.

The main results of this work are summarized as follows.
While the analyses are based on the specification of CRIS
onboard {\it EDGE} or {\it Xenia}, essentially the same
conclusions are 
applicable for {\it DIOS} or {\it ORIGIN}, which are equipped with
a spectrometer of slightly different FOV, angular resolution,
and effective area.

\begin{itemize}
	\item The WHIM can be detected in emission through  \ion{O}{7}  and \ion{O}{8} lines.
	Unambiguous detection of the line systems requires the simultaneous detection of
	both lines, each one with statistical significance above
	      5$\sigma$. With a 1 Ms observation 
	and in a  FOV of 1 deg$^2$ we expect 
	to detect $\sim 640$   \ion{O}{7} + \ion{O}{8} line emitting
	      systems.
A significant number of detections ($\sim 300$ per deg$^{2}$) 
	      is still expected with    
	a much shorter (100~ks) observation. These results are in broad agreement with those of
	\cite{2006ApJ...650..573C}.
	\item A significant number of these line systems, however, trace gas which is hotter or
	denser than the WHIM ($\delta>1000$).
	      The number of line systems contributed by the WHIM is $\sim 65$ \%
	(i.e., $\sim 430$ deg$^{-2}$) for a 1 Ms observation. The fraction drops to  $\sim 20$ \%
          when the exposure time is reduced to 100~ks. In other words, deep observations are required
          to detect a number of line systems large enough to investigate the statistical properties of the  
          WHIM. Shorter observations would preferentially probe hotter gas in high density environments   
          associated to clusters or groups.
          \item The expected large number of detections is robust. If one considers the 
          very conservative model B1, which we have implemented to assess theoretical
          uncertainties, the number of line systems above 5$\sigma$ detection threshold is still
          large ($\sim 480$ deg$^{-2}$), 40\% of which contributed by
		the WHIM. 
		The difference between model B2 and model B1 provides generous estimate of the theoretical
          uncertainties, significantly larger than the scatter among the models predictions of 
          \cite{2006ApJ...650..573C}.          
          \item 
		The number of detectable systems is affected by 
		contamination from 
          spurious lines and by the level of the continuum that may
		vary with energy and direction.
          It is confirmed that a more rigorous detection procedure that
		accounts  for both effects misses only
		$\sim 30$ \% of the line systems that are originally
		selected. 

 \item The number of detectable emission lines decreases with the energy 
       resolution. However, even with an energy resolution of 7~eV, 
       already achieved by the microcalorimeter on board {\it Suzaku}, 
       one expects to detect $\sim$160~ \ion{O}{7}$+$\ion{O}{8}
       line systems per deg$^2$ in~1 Ms observation.
 \item Detectable line systems preferentially  trace the denser regions of the WHIM with typical 
          gas overdensity of $\deltag \sim 100$, typically associated to virialized structures like galaxies,
          groups, and the outskirts of galaxy clusters. Increasing the instrumental sensitivity will not change 
          these results significantly since the bulk of the WHIM that resides in lower density environment
          is characterized by a very  low line emissivity.
          These results agree with those of  \cite{2010MNRAS.407..544B} and illustrates the importance 
          of studying the WHIM both in emission (to assess its spatial distribution and thermal state)
          and in absorption (to probe low density regions and close the baryon budget).
           \item In the region where CIE holds, the temperature of the gas associated to the detectable emission systems can be reliably 
           estimated from the \ion{O}{7}/\ion{O}{8} line ratio in the
		 range 
		 of (1--5)$\times 10^6$~K, with typical random 
           errors of $\sim 20$ \%. The measured temperature  trace the \ion{O}{7}-emission weighted temperature 
           of the  gas, which, for  $T=$(1--3)$\times 10^6$~K, probes the bulk of  the gas in the emitting region.
           \item With 400--600 line emission systems detected per square degree, one can 
           trace the spatial distribution of the line emitting gas, a large fraction of which is in the warm-hot    
           phase. The visual inspection of the emission maps confirms that the selected line systems 
           trace the large scale filamentary structure of the gas with typical overdensity  
           $\rho_\mathrm{gas}/\left<\rho_\mathrm{gas}\right>\ge75$. Since line emission is associated 
           to the denser part of the WHIM, the tracing is rather sparse and concentrated around the
           virialized structures that punctuate the filaments. Moreover, because of the unfavorable 
           ionization balance, the tracing is biased against the largest 
		 virialized structures such as galaxy
           clusters that are typically found at the nodes of the
		 network of filaments.

 \item The observed $dN/dz$ relation can be compared to theoretical
       predictions to discriminate among WHIM models and constrain the
       physical processes that regulate its thermal state and
       chemical composition. As an example we have shown that the
       $dN/dz$ statistics is capable of
       discriminating among different metallicity models with 
       a significance of
       3--5$\sigma$ with a 1~deg$^2$ observation, 
       depending on the energy resolution. 
       Further constraints will be provided by the analysis of the
       3D distribution of the line emission systems and by combining these
       results with those obtained from the analysis of the absorption spectra.
\end{itemize}
 
\acknowledgments

The authors would like to thank an anonymous referee for the important
comments that helped improving the manuscript.
The authors are grateful to Stefano Borgani for providing us  the outputs of 
\cite{2004MNRAS.348.1078B} 
hydrodynamical simulation
realized using the IBM-SP4 machine at the ``Consorzio
Interuniversitario del Nord-Est per il Calcolo Elettronico'' (CINECA),
with CPU time assigned thanks to an INAF-CINECA grant. 
YT appreciates the fruitful discussion with Hiroshi Yoshitake
on the SWCX.
EU acknowledges financial
contribution from contracts ASI-INAF I/088/06/0 WP 15300
and ASI-INAF I/088/06/0 TH-018.
MV is supported by a PRIN-INAF, a PRIN-MIUR, INFN-PD51 and,
the ERC-StG "cosmoIGM".

SRON is supported financially by NWO, the Netherlands Organization for
Scientific Research.  This work is also supported by
JSPS grant-in-aid (KAKENHI 20840051, 22111513).

\bibliographystyle{apj}
\bibliography{ms}

\end{document}